\documentclass[smallextended]{svjour3}       
\usepackage{short_commands_article}

\smartqed  
\usepackage{natbib}        
\usepackage{makecell}
\usepackage{booktabs}
\usepackage{lineno}
\usepackage{stackengine}
\usepackage{enumerate}
\usetikzlibrary{arrows}

\journalname{Journal of Discrete Event Dynamic Systems}
\begin{document}
\title{Incremental Observer Reduction Applied to Opacity Verification and Synthesis
\thanks{This work was partly carried out within the project SyTec – Systematic Testing of Cyber-Physical Systems, a Swedish Science Foundation grant for strong research environment. \\The support is gratefully acknowledged.}
\thanks{Some of the results in this paper were presented in preliminary form at the 
14th International Workshop on Discrete Event Systems (WODES), Sorrento Coast, Italy, May 2018 \citep{mona_18}.}}
\author{Mona Noori-Hosseini, Bengt Lennartson and Christoforos N. Hadjicostis}
\institute{M. Noori-Hosseini and Bengt Lennartson\at Division of Systems and Control, Department of Electrical Engineering \at Chalmers University of Technology, SE-412 96 G\"{o}teborg, Sweden \\  \email{noori@chalmers.se, bengt.lennartson@chalmers.se} \\
Christoforos N. Hadjicostis \at Department of Electrical and Computer Engineering \at University of Cyprus, Nicosia, Cyprus\\ \email{chadjic@ucy.ac.cy}}
\date{Received: date / Accepted: date}
\maketitle
\begin{abstract}
An incremental observer generation for modular systems is presented in this paper. It is applied to verification and enforcement of current-state opacity and current-state anonymity, both of which are security/privacy notions that have attracted attention recently. The complexity due to synchronization of subsystems, but also the exponential observer generation complexity, are tackled by  local observer generation and an incremental abstraction. Observable events are hidden and abstracted step by step when they become local after synchronization with other subsystems. For systems with shared unobservable events, complete observers can not be generated before some local models are synchronized. At the same time, observable events should be abstracted when they become local, to avoid state space explosion. Therefore, a new combined incremental abstraction and observer generation is proposed. This requires some precaution (detailed in the paper) to be able to accomplish local abstractions before shared unobservable events are removed by observer generation. Furthermore, it is shown how current state opacity and anonymity can be enforced by a supervisor. This is achieved by a natural extension of the verification problem to a supervisory control problem based on forbidden states and incremental abstraction. Finally, a modular and scalable building security problem with arbitrary number of floors and elevators is presented, for which the efficiency of the incremental abstraction is demonstrated.
\end{abstract}

\section{Introduction} \label{intro}
With the rapid growth of large communication networks and online services, and their diverse applications, ranging from modern technologies in defense and e-banking to health care and autonomous vehicles, security and privacy concerns on their information flow are raised. This means that unauthorized people should not acquire the information flow in these services, for instance in terms of being able to track and identify  real time location information about the users. There are various notions on security and privacy for different applications based on their vulnerability to intruders. Next we provide some background information and related literature on topics that are relevant to the developments in this paper.

\paragraph{Opacity verification} One category of security notions, \citep{Focardi} concerns the information flow from the system to the outside observer, which is called opacity \citep{Saboori_2007,lesage16}. Opacity is a general and formal security property that has been widely investigated for discrete event systems (DESs) for finite automata \citep{Saboori_2007,bryan08,Saboori_2008,Saboori:2014}, but also for {P}etri nets \citep{Bryans_2005,Tong_2016,tong:giua:17}.  A system is opaque if, for any secret behavior, there exists at least one non-secret behavior that looks indistinguishable to the intruder \citep{Saboori_2007,laf18}. The security notion is investigated for automata \citep{lesage16} using either state-based predicates \citep{Saboori_2007,hadji11,lesage16,Tong:giua:2017}, or language-based predicates \citep{Badouel:2007,Saboori_2008,Cassez_2009,Lin_2011,Tong_2016}. 

Depending on the modeling formalism of the system and the secret, there are different opacity notions, such as current-state opacity, initial-state opacity, and k-step opacity \citep{lesage16}. \cite{wu2013} show that there exists a polynomial-time transformation between different notions of opacity for finite automata and regular languages. Current-state opacity \citep{Saboori:2014,Tong:giua:2017} requires that the sequence of observable events seen by the intruder never allows the external observer to unambiguously determine that the current state of the system falls within a given set of secret states. A number of examples and applications are presented in \cite{hadji11}. A privacy notion that is adapted from  current-state opacity is proposed in \cite{bryan08,Lin_2011}. It is called anonymity, and in \cite{Wu_2014} it is used for location privacy and is called current-state anonymity. The servers that access the user's location information are then regarded as intruders. 

An intruder with partial observation can be modeled as an observer of the system, meaning that it has full knowledge about the system structure, while it is only able to see the observable events of the system. Observers achieved by subset construction \citep{cl:int:2008} are deterministic finite automata that estimate the set of possible current states for verifying properties of interest. There are several works that exploit observer generation for opacity verification \citep{Saboori_2011,Saboori_2013,wu2013,wu:abs:ver:2018}. 

\paragraph{Opacity enforcement}
Ensuring opacity on a system is usually performed by exploiting supervisory control \citep{rw:con:1989} as in \cite{Takai_2008} and \cite{Takai_2009}. Given a system that is not current-state opaque with respect to a secret, it is required to design a maximally permissive supervisor that restricts the behavior of the system to turn it into a current-state opaque system. The design of supervisors to enforce opacity is also sometimes called opacity enforcement. In \cite{Badouel:2007}, the language-based opacity and a set of intruders having different observations are considered. The work by \cite{dubreil08,dubreil10} is also focused on  language-based opacity enforcement for one intruder. 

Enforcing opacity using supervisory control techniques is also investigated by \cite{Saboori_2008}. They propose methods for designing optimal supervisors to enforce two different opacity properties, with the assumption that the supervisor can observe all controllable events \citep{saboori:2012}. In the work by \cite{Yin_2016,Tong:giua:2018}, to enforce current-state opacity, the assumption that all controllable events should be observable is relaxed.  In \cite{Wu_2014_automatica,Ji_2018} a novel enforcement mechanism is proposed, based on the use of insertion functions that change the output behavior of the system, by inserting additional observable events. 

\paragraph{Modularity and abstraction}
To verify or synthesize a supervisor to enforce  current-state opacity/anonymity in a modular system, it is required to generate the system's observer. Given the exponential complexity of observer generation, as well as the complexity of interacting subsystems, especially for large complex modular systems, state space explosion often occurs while performing verification or synthesis. For this reason, reduction methods play an important role in making the procedure feasible. \cite{HadjAlouane_2017} use a binary decision diagram technique \citep{Bryant_1992} to abstract graphs of moderate size, as a method for the verification of three different opacity variants. Moreover, they prove that  opacity properties are preserved by composition, which guarantees that local verification of these properties can also be performed. 

In \cite{Zhang_2017}, a bisimulation-based method to verify the infinite-step opacity of nondeterministic finite transition systems is proposed. Since this abstraction is based on strong bisimulation it has a minor reduction capability compared to abstractions where local events are hidden, such as weak bisimulation \citep{m:com:1989} and branching bisimulation \citep{glabbeek96}. Recently the authors have proposed an abstraction method for current-state opacity verification of modular systems \citep{mona_18} based on a similar abstraction, called visible bisimulation equivalence \citep{bl_18}. Both state labels and transition labels (events) are then integrated in the same abstraction method. This abstraction has the benefit that temporal logic properties are preserved in the abstraction, and the opacity verification in \citep{mona_18} is formulated as a temporal logic safety problem.

\paragraph{Incremental observer abstraction}
In the abstraction, local events (only included in one subsystem) are hidden and then abstracted such that temporal logic properties related to specific state labels are still preserved. When subsystems are synchronized more local events are obtained, which also means that more events can be hidden and abstracted. This hiding/abstraction method is repeated until all subsystems have been synchronized. The result is an incremental abstraction technique where state space explosion is avoided when a reasonable number of events are local or at least only shared with a restricted number of subsystems. Most real systems have this event structure, and still some events can be shared by all subsystems. This incremental abstraction technique for modular systems can be traced back to \citep{graf:96}, but its application to local events was more recently proposed in \citep{flordal09}, where it was called compositional verification. In \citep{mona_18} this incremental abstraction is adapted to opacity verification, and it shows great computational time improvement compared to standard methods.

\paragraph{Nonblocking transformation} In this paper, both current state opacity verification and current state anonymity verification are formulated based on state labels in transition systems. Non-safe states in corresponding local observers are then naturally considered as forbidden states. By introducing simple detector automata, the problem is easily transformed to a nonblocking problem. For this modular system, the efficient conflict equivalence abstraction in \cite{Malik04} and \cite{flordal09} is used, since it preserves the nonblocking properties of the original modular observer. The reason for evaluating this abstraction is that it is known to be more efficient than visible bisimulation. This abstraction has independently been proposed for opacity verification by \cite{sahar_opacity:2019} and \cite{mona_opacity:2019}. In both reports, the abstraction gives an enormous reduction in computation time, compared to opacity verification without abstraction. In our work, the procedure is evaluated on a scaleable building security problem, including an arbitrary number of floors and elevators. 

\paragraph{Observer abstraction including shared unobservable events}
Two main extensions are also included in this paper, first the nontrivial introduction of shared unobservable events. It means that complete local observers can not be computed before some local models are synchronized. The reason is that shared unobservable events can not be reduced in the observer generation before they have become local after synchronization. At the same time, observable events should be abstracted when they become local, to avoid state space explosion. The proposed solution is to extend the incremental abstraction with an incremental observer generation, such that a switch between abstraction and observer generation can be performed when subsystems are synchronized. This requires some precaution to be able to accomplish local abstractions before shared unobservable events are removed by the observer generation. Some minor restrictions are  included to be able to prove that the combined incremental observer generation and abstraction works correctly. This procedure includes additional temporary state labels, which motivates the more general and flexible visible bisimulation abstraction. 

\paragraph{Incremental supervisor abstraction for opacity enforcement}
To enforce opacity and anonymity it is also shown how an observer based maximally permissive supervisor can be generated by incremental abstraction. This supervisor generation follows naturally as an extension of the original forbidden state formulation of opacity and anonymity verification. The incremental abstraction is based on a supervision/synthesis equivalence proposed by \cite{fmf:sup:2007,sahar14,sahar17} as a natural extension of conflict equivalence \citep{Malik04}. 

\paragraph{Main contributions}
To summarize, the main contributions of this paper are: 1) a transition system based formulation of modular observers applied to current state opacity and current state anonymity verification, 2) a simple transformation of the modular observer verification problem to a nonblocking problem based on simple detector automata, a generic technique that can be applied to many verification and synthesis problems, for instance abstraction based diagnosability verification \citep{mona_diag:2019}, 3) a combined incremental observer generation and abstraction for modular systems including shared unobservable events, 4) an incremental abstraction based synthesis  of observer based maximally permissive supervisors for current state opacity and anonymity, 5) a modular formulation of a scaleable building security problem including an arbitrary number of floors and elevators, and finally 6) a demonstration on how efficient the proposed incremental abstraction of observers for current state opacity and current state anonymity verification and synthesis works for large modular systems.

The remainder of the paper is organized as follows. After some preliminaries introduced in Section 2, the problem statement is presented in Section 3. Efficient generation of modular observers is shown in Section 4, followed by some specific results on current-state opacity/anonymity for modular systems in Section 5.  In Section~6, it is shown how a combined incremental observer generation and abstraction can be achieved for systems including shared unobservable events. Section 7 presents a scaleable floor/elevator building for which the efficiency of the proposed incremental abstraction is demonstrated. In Section 8, an incremental abstraction based supervisor generation for current state opacity and anonymity is developed, followed by some concluding remarks in Section 9.

\section{Preliminaries}\label{sprel}
\vspace{-1ex}
A {\em transition system} $G$ is defined by a 6-tuple $G=\langle X,\Sigma,T, I,AP,\lambda \rangle$ where $X$ is a set of states, $\Sigma$ is a finite set of events, $T\subseteq X\times \Sigma\times X$ is a transition relation, where $t=(x,a,x')\in T$ includes the source state $x$, the event label $a$, and the target state $x'$ of the transition $t$. A transition $(x,a,x')$ is also denoted \mbox{$x\trans{a}x'$}. $I\subseteq X$ is a set of possible initial states, $AP$ is a set of atomic propositions, and \mbox{$\lambda : X \ra 2^{AP}$} is a state labeling function. 

A subset $\mc L \subseteq \Sigma^*$ is called a \emph{language}. Moreover, for the event set $\Omega \subseteq \Sigma$, the {\em natural projection}  $P:\Sigma^*\ra \Omega^*$ is  inductively defined as $P(\veps)=\veps$, $P(a) = a$ if $a\in\Omega$,  $P(a) = \veps$ if $a\in\Sigma\setm\Omega$, and $P(sa)=P(s)P(a)$ for $s\in\Sigma^*$ and $a\in \Sigma$. In the composition of subsystems, see \defr{dsynch}, events that are not included in any synchronization with other subsystems are called {\em local events}. Such events are central in the abstraction of observers.

\paragraph{Modeling $\veps$ transitions.} The transition system $G$ is now extended to include transitions labeled by the empty string $\veps$. In this paper, the $\veps$ label will explicitly be used for {\em local unobservable events}. If nothing special is pointed out, it means that such local unobservable events are replaced by $\veps$ and therefore not included in the alphabet $\Sigma$, while the total alphabet is extended to $\Sigma\cup \{\veps\}$. A  sequence of $\veps$ transitions $x=x_0\trans{\veps} x_1\trans{\veps} \cdots \trans{\veps} x_n=x'$, $n\geq 0$, is denoted $x\transd{\veps}x'$. A corresponding sequence, including possible $\veps$ transitions before, after and in between events in a string $s\in \Sigma^*$, is denoted $x\transd{s}x'$. The \mbox{\em epsilon closure} of a state $x$ is defined as $R_\veps(x)=\{x'\st x\transd{\veps} x'\}$, and for a set of states $Y\subseteq X$ we write $R_\veps(Y)=\bigcup_{x\in Y} R_\veps(x)$.

A nondeterministic transition system generally includes a set of initial states,  \text{$\veps$ labeled} transitions, and/or alternative transitions with the same event label. A transition function for an event $a\in\Sigma$ in a nondeterministic transition system is defined as $\delta(Y,a)=R_\veps(\{x'\st (\exists x \in R_\veps(Y)) \;x\trans{a}x'\in T\})$. An extended transition function is then inductively defined, for $s\in \Sigma^*$ and $a\in \Sigma$, as $\delta(I,sa)=\delta(\delta(I,s),a)$ with the base case $\delta(I,\veps)=R_\veps(I)$. Furthermore, the language for a nondeterministic transition system is  defined as $\mc L(G)=\{s \in \Sigma^* |(\exists x\in I) \, \delta(x,s)\neq \varnothing  \}$. 

\paragraph{Local transitions and hidden $\tau$ events}
To obtain efficient abstractions, a special $\tau$ event label is used for transitions with local observable events. The lack of communication with other subsystems means that the $\tau$ event is hidden from the rest of the environment. The closure of $\tau$-transitions in a finite path $x=x_0 \trans{\tau}x_1\trans{\tau} \cdots\trans{\tau}x_n=x'$, $n\geq 0$ is denoted \mbox{$x\transd{\tau}x'$}. 

Note the difference between $\veps$ and $\tau$ events. Unobservable local events are replaced by $\veps$ before an observer is generated, which  removes any $\veps$ transitions. Observable local events are then replaced by $\tau$ to model that they are hidden before performing any abstraction. In process algebra, the replacement of any specific event by the event $\tau$ is called {\em hiding}, cf. \cite{m:com:1989}.  A transition system $G$ where the events in $\Sigma^h$ are hidden and replaced by~$\tau$ is denoted~$G^{\Sigma^h}$.

\paragraph{Partition $\Pi$ and block $\Pi(x)$} 
To obtain abstracted transition systems, states $x,y\in X$ that can be considered to be equivalent in some sense, denoted \mbox{$x\sim y$}, are merged into equivalence classes \mbox{$\ec{x}=\{y\in X\st x\sim y\}$}, also called blocks. These blocks, which are non-overlapping subsets of~$X$, divide the state space into the {\em quotient set}  \mbox{$X/\mathord{\sim}$}, also called a partition $\Pi$ of $X$. The block/equivalence class including state~$x$ is  denoted \mbox{$\Pi(x)=\ec{x}$}.  A partition $\Pi_1$ that is {\em finer} than a partition $\Pi_2$, denoted $\Pi_1\preceq\Pi_2$, means that $\Pi_1(x)\subseteq \Pi_2(x)$ for all $x\in X$. The partition $\Pi_2$ is then said to be {\em coarser} than $\Pi_1$.

\paragraph{Invisible, visible and stuttering transitions}
For a given state partition~$\Pi$, a transition \mbox{$x\trans{\tau}x'$} is invisible if \mbox{$\Pi(x) =\Pi(x')$}, while a transition \mbox{$x\trans{a}x'$} is visible if \mbox{$a\neq\tau$} or \mbox{$\Pi(x)\neq\Pi(x')$}. A path \mbox{$x\trans{\tau}x_1\trans{\tau}\cdots\trans{\tau} x_n\trans{a}x'$} is called a {\em stuttering transition}, denoted \mbox{$x\transtau{a}x'$}, if \mbox{$\Pi(x)=\Pi(x_1)$} \mbox{$=\ldots = \Pi(x_n)$}, and \mbox{$a\neq\tau$} or $\Pi(x_n)\neq$ $\Pi(x')$. This means that the first $n$ transitions are invisible, while the last one is visible. A {\em block stuttering transition} corresponding to \mbox{$x\transtau{a}x'$} is denoted \mbox{$\Pi(x)\transtau{a}\Pi(x')$}.

\paragraph{Visible bisimulation} Different types of bisimulations, used for abstraction, are either defined for labeled transition systems, only including event labels (often called actions) on the transitions, or for Kripke structures, only including state labels \citep{baier08}. In this work, shared events are required for synchronization of subsystems, while  state labels are used to model security properties. Recently,  \cite{bl_18} introduced an abstraction for transition systems including both event and state labels, called visible bisimulation. It is directly defined as an equivalence relation based on block stuttering transitions, and more specifically  on the set of event-target-blocks $\gmb(x) =  \{\transtau{a}\Pi(x') \st x\transtau{a}x'\}$ that defines all possible stuttering transitions from an arbitrary state $x$.

\begin{definition}[Visible bisimulation equivalence]\label{dbs}
{\rm
Given a transition system $G=\langle X,$ $\Sigma,T,I,AP, \lambda \rangle$ and the state label partition  $\Pi_\lambda(x)=\{y\in X\st \lambda(x)=\lambda(y) \}$, 
a partition $\Pi$, for all $x\in X$ determined by the greatest fixpoint of the fixpoint equation
\[
\Pi(x) = \{y\in X \st  \Pi\preceq\Pi_\lambda\AND \gmb(x) = \gmb(y) \},
\]
is a {\em visible bisimulation (VB) equivalence}, and states $x,y\in \Pi(x)$ are visibly bisimilar, denoted $x\sim y$. \ebox
}\end{definition}

\paragraph{Quotient transition system}
Blocks are the states in abstracted transition systems, and the notion partition $\Pi$ is used in the computation of this model, while the resulting reduced model takes the equivalence  perspective.  It is therefore called  {\em quotient transition system}, and for a given partition $\Pi$ it is defined as $G\qsb=\langle X\qsb,\Sigma,T\qssb,$ $I\qssb,AP,\lambda\qssb \rangle$, where \mbox{$X\qsb=\{\ec{x} \st \ec{x}=\Pi(x) \}$} is the set of block states (equivalence classes), \mbox{$T\qssb= \{ \ec{x}\trans{a} \ec{x'} \st  x \trans{a} x' ( \ec{x'}\neq \ec{x} \OR a\neq\tau)\big) \} $} is the set of block transitions, here specifically defined for VB, \mbox{$I\qssb= \{\ec{x} \st x\in I \}$} is the set of initial block states, and \mbox{$\lambda\qssb(\ec{x})= \lambda(x)$} is the block state label function, where it is assumed that \mbox{$\lambda(x)= \lambda(y)$}, $\forall y \in \ec{x}$. 

Visibly bisimilar states $x\sim y$ in $G$ are also visibly bisimilar to the block state~$\ec{x}$ in $G\qsb$, \ie $\ec{x}\sim x$ for all $x\in X$. Furthermore, $G$ and $G\qsb$ are VB equivalent, denoted $G\sim G\qsb$. Combining hiding of a set of events $\Sigma^h$ for a system $G$, followed by the generation of the quotient transition system, results in the {\em abstracted transition system} $G^{\Sigma^h}\!\smm \qsb \define G^{\mathcal A^{\Sigma^h}}$\!. This also means that $G^{\Sigma^h}\!\smm\sim G^{\mathcal A^{\Sigma^h}}$\!.

\paragraph{Synchronous composition}
The definition of the synchronous composition in \cite{h:com:1985} is adapted to $\tau$ events, where such events in different subsystems are not synchronized, although they share the same event label. They are simply considered as local events, which is natural since the hiding mechanism where an event is replaced by the invisible $\tau$ event is only applied to local events. This results in the following definition of the synchronous composition, including $\tau$ event labels.

\begin{definition}[Synchronous composition including $\tau$ events] \label{dsynch}  \xspa{0.1em}
{\rm
Consider two transition systems $G_i=\langle X_i,\Sigma_i,T_i,I_i,AP_i, \lambda_i \rangle$, $i=1,2$. The synchronous composition of $G_1$ and $G_2$ is defined as
\[
G_1 \synch G_2=\left\langle X_1\times X_2,\Sigma_1 \cup \Sigma_2, T , I_1 \times I_2 , AP_1 \cup AP_2, \lambda \right\rangle
\]
where
{\fontsize{9.5}{9.5}\selectfont
\[ \begin{array}{ll}
  (x_1,x_2)\trans{a} (x'_1,x'_2)\in T: & a \in (\Sigma _1 \cap \Sigma _2)\setminus \{\tau\},\; x_1\trans{a}x'_1\in T_1,\; x_2\trans{a}x'_2\in T_2, \yspa{1.3ex} \\
  (x_1,x_2)\trans{a} (x'_1,x_2)\in T: & a \in  (\Sigma _1 \smm\setminus\smm \Sigma _2 )\cup\{\tau\}, \; x_1\trans{a}x'_1\in T_1, \yspa{1.3ex} \\
  (x_1,x_2)\trans{a} (x_1,x'_2)\in T: & a \in  (\Sigma _2 \smm\setminus\smm \Sigma _1 )\cup\{\tau\}, \; x_2\trans{a}x'_2\in T_2,
\end{array} \]}
and
$\lambda : {X_1} \times {X_2} \to 2^{AP_1 \cup AP_2}.$ \hfill \ebox}
\end{definition}
Any transitions with $\veps$ labels, representing local unobservable events, are handled in the same way as  $\tau$ event labels, representing observable local events, since both stand for local events. On the other hand, before subsystems are synchronized, local observers will in this work be generated. This means that any $\veps$  transitions will be removed before synchronization.

\paragraph{Nonblocking and controllable supervisor}
In order to determine whether a system satisfies a given specification or not, the system has to be {\em verified}, and if it fails, the system is restricted by \emph{synthesizing} a \emph{supervisor}.  This means that states from which it is not possible to reach a desired marked state, called \emph{blocking states}, are removed. Furthermore, any uncontrollable events that can be executed by the plant are not allowed to be disabled by the supervisor \citep{rw:con:1989,Wonham_2017}.  Thus, a supervisor is synthesized to avoid blocking states and disabling uncontrollable events. Such a nonblocking and controllable supervisor is also maximally permissive, meaning that it restricts the system as little as possible.

\section{Problem statement}\label{prob_f}
The focus of this paper is to generate reduced observers that still preserve relevant properties, to be able to verify different security notions. It is also shown how supervisors can be generated, avoiding states that do not satisfy desired properties. This section presents the main problem statements of the paper, the incremental generation of reduced observers, and some security notions that will be analyzed by such reduced observers. First observers only involving local unobservable events are considered, where all such local events are immediately replaced by $\veps$. The more complex case, where some unobservable events are shared between different subsystems, means on the other hand that the shared unobservable events can not be replaced by $\veps$ before they have become local due to synchronization.

\subsection{Incremental abstraction for modular systems}
A transition system, including a number of subsystems $G_i$, $i\in\mathbb{N}^+_n$ that are interacting by  synchronous composition, is defined as 
\begin{equation}\label{modular}
G\sa= \,\saa\parallel_{i\in\mathbb{N}^+_n}\smm G_i=G_1\synch G_2\synch\cdots\synch G_n.
\end{equation}
A straightforward approach to analyze such a modular system is to compute the explicit monolithic transition system $G$. However, there are limitations on memory and computation time in the generation and analysis of such monolithic systems. An alternative approach is to avoid building the explicit monolithic system, by analyzing each individual subsystem first. In this case, local events of each subsystem are hidden and abstracted based on the desired property to be preserved. Moreover, after every synchronization of subsystems more local events may appear and thus, additional abstraction is possible. This step by step combined hiding, abstraction and synchronization is here called \emph{incremental abstraction}. In \citep{flordal09}, this approach is proposed for verification, and is called \emph{compositional verification}.

\subsection{Incremental observer generation including abstraction}\label{incobs}
The focus of this paper is on verification of security properties, while a simple extension towards synthesis is shown in the end of the paper. The security properties are analyzed by constructing an observer, where only observable events are involved. The generated observer is deterministic and computed by subset construction \citep{hmu:int:2001}. 

Since the observer generation as well as the synchronization of the subsystems have exponential complexity, the incremental abstraction mentioned above is of interest. This approach can be applied if the observer generation is divided into local observers that are synchronized. When all unobservable events are local, \ie no shared unobservable events are involved, it is shown in Section~\ref{obs:gen} that an observer of the monolithic system $G$, denoted $\mathcal O(G)$, also can be computed by the synchronous composition of the local observers of its subsystems. Thus,
\begin{equation}\label{obs_eq}
\mathcal O(G) = \,\saa\parallel_{i\in\mathbb{N}^+_n}\smm \mathcal O(G_i).
\end{equation}
The security properties considered in this work result in observer states that are either safe or non-safe. Introducing the state label $N$ for the non-safe states, visible bisimulation can be used in an incremental abstraction, still preserving the separation between the two types of states. 

For two synchronized subsystems, $G_1\synch G_2$, the sets of local events in $G_1$ and $G_2$ are $\Sigma^h_1$ and $\Sigma^h_2$, respectively, and the events in $\Sigma^{h}_{12}$ are the shared events between the two subsystems that become local after the synchronization, see also \exr{ex_sig}. Thus, the set $\Sigma^{h}= \Sigma^{h}_1 \saa\dot\cup \Sigma^{h}_2 \saa\dot\cup \Sigma^{h}_{12}$ includes all events that can be hidden after the synchronization. Using the notations $G^{\Sigma^h}$ for hiding the events in $\Sigma^h$, $G^{\mathcal A^{\Sigma^h}}$  for abstraction including hiding, and the equivalence $G^{\Sigma^h}\!\smm\sim G^{\mathcal A^{\Sigma^h}}$\!, it is also shown in Section \ref{obs:gen} that an abstraction of $\mathcal O(G_1 \synch G_2)^{\Sigma^h} $, including the local observer generation in \rf{obs_eq}, can be incrementally generated as

\begin{equation}\label{abs_eq_old}
\mathcal O(G_1 \synch G_2)^{\Sigma^h}  \!
\sim  \big (\mathcal O(G_1)^{\mathcal A^{\Sigma^h_1}} \smm \synch \mathcal O(G_2)^{\mathcal A^{\Sigma^h_2}} \big )^{\mathcal A^{\Sigma^h_{12}}}.
\end{equation}
Repeating this incremental abstraction procedure when more subsystems are included still implies that only observers of individual subsystems $ \mathcal O(G_i)$ are required. Furthermore, the repeated abstraction means that often systems with a moderate state space are synchronized, especially when a number of local events are obtained after each synchronization.

Since \rf{abs_eq_old} only includes one type of state label ($N$), it can also be expressed in terms of marked and non-marked states. Therefore, the problem can also be identified as a non-blocking problem, and more efficient abstractions (coarser state partitioning) than visible bisimulation can be used. This is  further described in Section~\ref{obs:gen}. 

When no explicit set of hidden events is included in the abstraction operator $\mc A$, the default set of events to be hidden is assumed to be all local observable events. Assuming that this set is $\Sigma^h$ for transition system $G$, it means that  $\mc O(G)^{\mathcal A^{\Sigma^h}}$ is often simplified to  $\mc O(G)^{\mathcal A}$, where we also note that the observer is generated before the abstraction is performed.

\subsection{Incremental observer generation with shared unobservable events} \label{subsec:abs}
For systems also including shared unobservable events, such events can not be replaced by $\veps$ due to the synchronization with other subsystems. This means that a complete observer can not be computed by composing local observers as in \rf{obs_eq} before subsystems have been synchronized such that no shared unobservable events remain. On the other hand it is shown in Section \ref{mixed} that observers can also be computed incrementally, such that shared unobservable events have to be retained, while transitions with local unobservable events can be removed in a partial observer generation.

To clarify this partial  observer generation, the more detailed observer operator $\mathcal O_{\Sigma^\veps}(G)$ is introduced, where the subscript $\Sigma^{\veps}$ includes the set of local unobservable events that are replaced by $\veps$ before the observer generation.  Similar to the sets of hidden events in \rf{abs_eq_old}, the sets of local unobservable events in~$G_1$ and $ G_2$ are $\Sigma^{\veps}_1$ and $\Sigma^{\veps}_2$, respectively, and the events in the set $\Sigma^{\veps}_{12}$ are the shared unobservable events in~$G_1$ and $G_2$  that become local after the synchronization $G_1\synch G_2$, see also \exr{ex_sig}. Thus, the set $\Sigma^{\veps}= \Sigma^{\veps}_1 \saa\dot\cup  \Sigma^{\veps}_2 \saa\dot\cup  \Sigma^{\veps}_{12}$ includes all unobservable events that can be replaced by~$\veps$ when the observer is generated after the synchronization. In Section~\ref{mixed} it is shown that an observer alternatively can be generated incrementally as

\begin{equation}\label{obs_eq_s}
\mathcal O_{\Sigma^{\veps}}(G_1\synch G_2) = \mathcal O_{\Sigma^{\veps}_{12}}\big ( \mathcal O_{\Sigma^\veps_1}(G_1)\synch \mathcal O_{\Sigma^\veps_2}(G_2)\big ),
\end{equation}
where the shared unobservable events in  $\Sigma^{\veps}_{12}$ are preserved until they become local. Also observe the special case with no shared unobservable event ($\Sigma^{\veps}_{12}=\varnothing$), where (\ref{obs_eq_s}) simplifies to (\ref{obs_eq}). Furthermore, the observer generation, combined with the incremental abstraction, results in the equivalence

\begin{equation}\label{abs_eq}
\mathcal O_{\Sigma^{\veps}}(G_1 \synch G_2)^{\Sigma^h} \!\sim \mathcal O_{\Sigma^{\veps}_{12}}\big (\mathcal O_{\Sigma^{\veps}_1}(G_1)^{\mathcal A^{\Sigma^h_1}} \smm\synch \mathcal O_{\Sigma^{\veps}_2}(G_2)^{\mathcal A^{\Sigma^h_2}} \big )^{\mathcal A^{\Sigma^h_{12}}}.
\end{equation}
Note that the observer generation is always performed before corresponding abstraction.  Observable and unobservable events are here incrementally replaced by $\tau$ and~$\veps$, respectively, when they become local. The mix between step-wise abstraction and partial observer generation means that some events are replaced by $\veps$ first after one or more abstractions. To be able to construct correct partial observers, this implies that some restrictions must be included in the incremental abstractions. This is solved in Section~\ref{mixed} by introducing  additional temporary state labels (other than labels for non-safe states). 

When no explicit set of unobservable events is included in the observer operator~$\mc O$, the default set is assumed to be all local unobservable events. Assuming that this set is $\Sigma^\veps$ for transition system $G$, it implies that  ${\mc O}_{\Sigma^\veps}(G)$ is often simplified to  $\mc O(G)$, where the observer is generated after the events in $\Sigma^\veps$ have been replaced by~$\veps$.

\b{example}\label{ex_sig}
This example illustrates the incremental replacement of local events by~$\veps$ or $\tau$ in \rf{abs_eq}. The events $a$, $b$, $c$ and $d$ are observable, while the events $u$ and $v$ are unobservable.  \figr{sig} shows that the events $a$, $d$ and $v$ are shared. To generate the local observers $\mathcal O(G_i)$, $i=1,\dots,3$, local unobservable events are replaced by $\veps$, and $\Sigma^{\veps}_1=\{u\}$, $\Sigma^{\veps}_2=\varnothing$, and $\Sigma^{\veps}_3=\varnothing$. Although event $v$ is unobservable, it is shared between $G_1$ and $G_3$ and is not replaced by $\veps$ at this level. However, it becomes local after the synchronization $G_1 \synch G_3$, which means $\Sigma^{\veps}_{13}=\{v\}$. Moreover, $\Sigma^{\veps}_{12}=\varnothing$ and $\Sigma^{\veps}_{23}=\varnothing$. In the hiding process of local observable events before abstraction, the sets of hidden events are $\Sigma^{h}_1=\{b\}$, $\Sigma^{h}_2=\{c\}$, $\Sigma^{h}_3=\varnothing$, $\Sigma^{h}_{12}=\{a\}$, $\Sigma^{h}_{13}=\varnothing$, and $\Sigma^{h}_{23}=\{d\}$.
\ebox\e{example}

\fig{h}{   
\begin{tikzpicture}[->,xscale=1, yscale=1, automaton]
            \node at (-2.5em,1.4em) {$G_1$};
            \node[initial, state] (1) at (0,0) {$0$};
            \node[state] (2) at (4em,0em)    {$1$};
            \node[state] (3) at (8em,0em)    {$2$};
            \node[state] (4) at (12em,2em)    {$3$};
            \node[state] (5) at (12em,-2em)   {$4$};
            \path[->]
              (1) edge [] node [above] {$a$} (2)
              (2) edge [] node [above] {$b$} (3)
              (3) edge [] node [above] {$u$} (4)
              (3) edge [] node [above] {$v$} (5);
\end{tikzpicture}

\begin{tikzpicture}[xscale=1, yscale=1, automaton]
            \node at (-2.5em,1.4em) {$G_2$};
            \node[initial, state] (1) at (0,0) {$0$};
            \node[state] (2) at (4em,0em)    {$1$};
            \node[state] (3) at (8em,0em)    {$2$};
            \node[state] (4) at (12em,0em)    {$3$};
            \path[->]
              (1) edge [] node [above] {$a$} (2)
              (2) edge [] node [above] {$c$} (3)
              (3) edge [] node [above] {$d$} (4);
\end{tikzpicture}
\begin{tikzpicture}[xscale=1, yscale=1, automaton]
\node at (-3.5em,2.05em) {\xspa{2em}\yspatb{1em}{0em}$G_3$};
            \node[initial, state] (1) at (0,0) {$0$};
            \node[state] (2) at (4em,0em)    {$1$};
            \node[state] (3) at (8em,0em)    {$2$};
            \path[->]
              (1) edge [] node [above] {$d$} (2)
              (2) edge [] node [above] {$v$} (3);
\end{tikzpicture}
\vspace{0.3em}
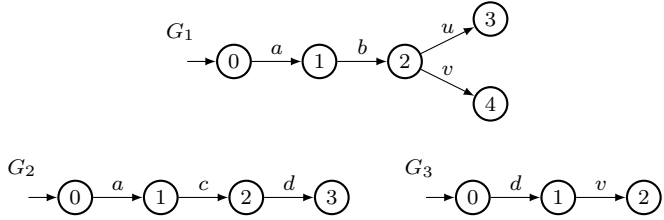
}{Three subsystems with local and shared, observable and unobservable events.}{sig}

\subsection{Opacity and privacy}\label{ssecprop}
The two security and privacy properties that are studied in this work are current-state opacity (CSO) and current-state anonymity (CSA). It is assumed that an intruder knows the model of the system and has access to the observable events. Thus, an intruder can generate an observer of the system, and security and privacy violation can  be formulated as the existence of non-safe states in this observer.

In CSO verification, the states of the observer that exclusively include secret states are called \emph{non-safe} states. By definition, a system is \emph{current-state opaque}, if there is no non-safe state in the observer. On the other hand, a system is \emph{current-state anonymous}, if there is no singleton state in the observer. The singleton states are considered as non-safe states in CSA verification. In Section \ref{op_an}, both opacity and anonymity notions for modular systems are described. 

Moreover, for the synthesis of current-state opaque/anonymous systems, that is limited to systems including only local unobservable events, uncontrollable events are introduced such that an efficient supervision equivalence abstraction can be used to find the supervisor.

\section{Efficient generation of observers} \label{obs:gen}
Since the computation of an observer has exponential complexity \citep{cl:int:2008}, it is shown in this section how the incremental abstraction in \rf{abs_eq_old} can be used to significantly lower the computational complexity. All unobservable events are in this section assumed to be local and can therefore immediately be replaced by~$\veps$. Based on this assumption, it is shown how local observers can be directly generated before the incremental abstraction is applied.

\subsection{Incremental observer abstraction for modular systems} \label{obs:gen:sub}
For a nondeterministic transition system $G$, where unobservable (local) events have been replaced by $\veps$, a deterministic transition system with the same language as $\mc L(G)$, called an \emph{observer} $\mathcal O (G)$, is  generated by subset construction \citep{hmu:int:2001}, where $\mathcal O (G) = \langle \what X,\Sigma,\what T,\what  I,AP,\what \lambda \rangle$, and $\what X=\{Y\in 2^X\st$ $ (\exists s\in \mc L(G)) \,Y=\delta(I,s)\}$, $\what T=  \{ Y\trans{a} Y' \st $ \mbox{$ Y'=\delta(Y,a) \} $}, and \mbox{$\what I=R_\veps(I)$}. The relation between $\what \lambda (Y)$ and $\lambda(x)$ is application dependent, see Section~\ref{op_an}, but the default assumption is that $\what \lambda (Y)=\bigcup_{x\in Y}\lambda(x)$. An obvious alternative is $\what \lambda (Y)=\bigcap_{x\in Y}\lambda(x)$, an interpretation that is applied in CSO.

Introduce the transition function $\what \delta(Y,a)\define\delta(Y,a)$ and the extended transition function, inductively defined as  $\what \delta(\what I,sa)=\what\delta(\what\delta(\what I,s),a)$ with the base case $\what \delta(\what I,\veps)$ $=\what I$. It is then easily shown that $\what \delta(\what I,s)=\delta(I,s)$, see~\citep{hmu:int:2001}. This means that $\mc L(\mathcal O(G))=\mc L(G)$.

For a modular system \rf{modular} with partial observation and no shared unobservable events, the monolithic observer can be computed by first generating local observers for each subsystem before they are synchronized. This is possible, since the same monolithic observer is obtained when synchronization is made before and after observer generation.  This was shown for automata by \cite{fabre:2012} and \cite{Pola_2017}. A minor extension to transition systems is presented in the following lemma. The first automata related part of the proof is included due to its simplicity compared to earlier formulations.

\begin{lemma}[Modular observers]\label{tobs}
{\rm Let $G_i=\langle X_i,\Sigma_i,T_i,I_i,AP_i, \lambda_i \rangle$, $i=1,2$, be two nondeterministic transition systems with no shared unobservable events,
where the alphabet $\Sigma_i$ only includes observable events. Then, the observer for the synchronized system}
\[
  {\mathcal O}{(G_1\synch G_2)} =  \mathcal O(G_1) \synch \mathcal O(G_2).
\]

{Proof:} {\rm  Consider the language of the synchronized system $\mc L(G_1\synch G_2)$ and the projection $P_i:(\Sigma_1\cup \Sigma_2)^* \ra \Sigma^*_i$ for $i=1,2$. After a string $s\in\mc L(G_1\synch G_2)$ has been executed, the set of reachable states can be expressed as $Y_1\times Y_2$, where
$
Y_i=\{x\st (\exists x_0\in I_i)\,x_0\stackrel{P_i(s)\yspa{0.8ex}}{\Longrightarrow} x\}, \;i=1,2.
$
Assume that there are transitions $x_i\transd{a}x'_i$ in $G_i$ for $i=1,2$, where $a\in \Sigma_1\cap \Sigma_2$, $x_i\in Y_i$, and $x'_i\in Y'_i$. Then there is a corresponding transition $(x_1,x_2)\transd{a}(x'_1,x'_2)$ in $G_1\synch G_2$. Thus,  subset construction of $G_1\synch G_2$ generates the transition $Y_1\times Y_2\trans{a}Y'_1\times Y'_2$. Since $Y_i$ and $Y'_i$ are also states in $\mathcal O (G_i)$, the corresponding transition in $\mathcal O (G_1) \synch \mathcal O (G_2)$ is $(Y_1, Y_2)\trans{a}(Y'_1,Y'_2)$.

With similar arguments for $a\in \Sigma_1\setminus \Sigma_2$ and $a\in \Sigma_2\setminus \Sigma_1$, we find that for a given string $s\in\mc L(\mathcal O {(G_1 \synch G_2))}= \mc L(\mathcal O (G_1) \synch \mathcal O (G_2))$, the reachable states included in the block states of $\mathcal O {(G_1 \synch G_2)}$ and $\mathcal O (G_1) \synch \mathcal O (G_2)$ are the same. Indeed, the bijective function 
$
f:2^{X_1\times X_2}\rightarrow 2^{X_1}\times 2^{X_2}, \txt{0.8}{where}  f(Y_1\times Y_2)=(Y_1,Y_2)
$
for $Y_i\in 2^{X_i},  \;i=1,2$, shows that the states in the two transition systems are isomorphic. The states and transitions are therefore structurally equal.

In \defr{dsynch}, the union of the state labels is taken in the synchronization. Together with the default assumption on union of state labels in observer block states, the state label of the synchronized block state $(Y_1,Y_2)=f(Y_1\times Y_2)$  becomes $\bigcup_{x_1\in Y_1}\lambda(x_1)\,\cup\, \bigcup_{x_2\in Y_2}\lambda(x_2)$. The alternative interpretation for CSO, where union is replaced by intersection in the observer generation, gives $\bigcap_{x_1\in Y_1}\lambda(x_1)\,\cup\, \bigcap_{x_2\in Y_2}\lambda(x_2)=\bigcap_{x_1\in Y_1}$ $\bigcap_{x_2\in Y_2}\big(\lambda(x_1)\,\cup\, \lambda(x_2)\big )$. The second formulation corresponds to synchronization before observer generation. The interpretation for CSA is shown in \secr{op_an}.
\ebox}\end{lemma}
\paragraph{Online estimation}  This lemma also has implications on online estimation of a modular system. Clearly, online estimation can be implemented by running local observers combined with online synchronization. Alternatively, one can simply maintain local sets of consistent estimates, which get synchronised when necessary (the latter approach avoids building and storing the local observers ahead of time, by essentially exploring only the observer states that are visited due to the particular sequence of observations that is seen). For either approach, the lemma results in a dramatic simplification on the complexity of online estimation.

In the following proposition, abstraction is added to the result of Lemma \ref{tobs}. The proposition is valid for any abstraction that is congruent with respect to (wrt) synchronization and hiding. The basic idea behind this incremental abstraction can be traced back to \cite{Malik04} and \cite{flordal09}.
\begin{proposition}[Incremental abstraction of modular observers]\label{prop:obs} 
{\rm Let $G_1$ and $G_2$ be two nondeterministic transition systems with no shared unobservable events but hidden observable events in the set $\Sigma^h\define\Sigma^{h}_1 \saa\dot\cup \Sigma^{h}_2 \saa\dot\cup \Sigma^{h}_{12}$, where  $\Sigma^h_i$ includes local events in $G_i$, $i=1,2$, and $\Sigma^{h}_{12}$ includes shared events in $G_1$ and $G_2$. For an arbitrary abstraction equivalence $G^{\Sigma^h}\!\smm\sim G^{\mathcal A^{\Sigma^h}}$ that is congruent wrt synchronization and hiding, the abstraction of the following observer can be incrementally generated as}
\[
\mathcal O(G_1 \synch G_2)^{\Sigma^h}  \!
\sim  \big (\mathcal O(G_1)^{\mathcal A^{\Sigma^h_1}} \smm \synch \mathcal O(G_2)^{\mathcal A^{\Sigma^h_2}} \big )^{\mathcal A^{\Sigma^h_{12}}}.
\]

{Proof:} {\rm Combining Lemma~\ref{tobs} with hiding of the local observable events in $G_1$ and $G_2$, we find that
$\mathcal O(G_1 \synch G_2)^{\Sigma_1^h\saa\dot\cup \Sigma^{h}_2} = 
\big(\mathcal O(G_1) \synch \mathcal O(G_2)\big)^{\Sigma_1^h\saa\dot\cup \Sigma^{h}_2}=
\mathcal O(G_1)^{\Sigma^h_1} \synch \mathcal O(G_2)^{\Sigma^h_2}.$
For an arbitrary equivalence $G\sim H$, congruence wrt synchronization means that $G\synch R \sim H\synch R$. Thus,
$
\mathcal O(G_1 \synch G_2)^{\Sigma_1^h\saa\dot\cup \Sigma^{h}_2} \sim 
\mathcal O(G_1)^{\mathcal A^{\Sigma^h_1}} \smm \synch \mathcal O(G_2)^{\Sigma^h_2} \sim
\mathcal O(G_1)^{\mathcal A^{\Sigma^h_1}} \smm \synch \mathcal O(G_2)^{\mathcal A^{\Sigma^h_2}}.
$
Now, also hiding the shared events in $\Sigma^h_{12}$, combined with congruence wrt hiding ($G\sim H$ implies $G^{\Sigma^h} \sim H^{\Sigma^h}$) and one more abstraction, we finally obtain
$
\mathcal O(G_1 \synch G_2)^{\Sigma_1^h\saa\dot\cup \Sigma^{h}_2\saa\dot\cup \Sigma^{h}_{12}} \sim 
 \big (\mathcal O(G_1)^{\mathcal A^{\Sigma^h_1}} \smm \synch \mathcal O(G_2)^{\mathcal A^{\Sigma^h_2}} \big )^{\mathcal A^{\Sigma^h_{12}}}.
$
\ebox}
\end{proposition}

\subsection{Incremental observer algorithm} \label{handling}
Based on \pror{prop:obs},  an incremental observer generation including abstraction is presented in Algorithm~1 for modular systems without any shared unobservable events. 
 
\fig{h}{
\begin{tabular}{  r l  }    
    \hline
    & \spa{-2.8ex}{\bf Algorithm 1}\;\;Incremental observer generation  including abstraction\\
    \hline
        & \spa{-0.1em} {\bf input} $G_1,\ldots,G_n$  \yspatb{0.75ex}{1.2ex}\\
        & \spa{-0.1em} {\bf output} $\mathcal O(G)^{\mathcal A}$ \yspa{1.2ex}  \\
    1: & \spa{-0.1em} {\bf for} $i\in \mathbb{N}^+_n$ {\bf do}   \yspa{1.2ex}  \\
    2: & \spa{1.3em} $G_{\{i\}}:=\mathcal O(G_i)$ \yspa{1.2ex}  \\
    3: & \spa{-0.1em} {\bf end for}   \yspa{1.2ex}  \\
    4: & \spa{-0.1em} $\pi_\Omega:=\{\{1\},\{2\},\ldots,\{n\}\}$  \yspa{1.2ex}  \\
    5: & \spa{-0.1em} {\bf repeat}             \yspa{1.2ex} \\
    6: & \spa{1.3em} Choose $\Omega_1,\Omega_2\in \pi_\Omega$ according to some heuristics \yspa{1.2ex}  \\
    7: & \spa{1.3em} $\Omega:=\Omega_1 \cup \Omega_2$           \yspa{1.2ex}  \\
    8: & \spa{1.3em} $G_\Omega:=G_{\Omega_1}^{\mathcal A} \synch G_{\Omega_2}^{\mathcal A}$  \yspa{1.2ex}  \\
    9: & \spa{1.3em} Replace $\Omega_1$ and $\Omega_2$ by $\Omega$ in $\pi_\Omega$ \yspa{1.2ex} \\
    10: & \spa{-0.1em} {\bf until} $\Omega=\mathbb{N}^+_n$ \yspa{1.2ex}  \\
    11: & \spa{-0.1em} $\mathcal O(G)^{\mathcal A}:=G_\Omega^{\mathcal A}$ \yspa{1.2ex}  \\
    \hline  		
\end{tabular}
\vspace{0.3em}
}{Observer generation and incremental abstraction of a modular transition system $G\sa= \,\saa\parallel_{i\in\mathbb{N}^+_n}\smm G_i$ without any shared unobservable events.}{alg:compos}
\paragraph{Heuristics} In the selection of the sets $\Omega_1$ and $\Omega_2$  and corresponding transition systems $G_{\Omega_1}$ and $G_{\Omega_2}$, to be abstracted in Algorithm 1, a natural approach is to first select a group of transition systems with few transitions. Among them, the two systems with the highest proportion of local events are chosen to be abstracted. In this way, a significant reduction of states and transitions is achieved by the abstractions, and the intermediate system after the synchronization $G_\Omega:=G_{\Omega_1}^{\mathcal A} \synch G_{\Omega_2}^{\mathcal A}$ also becomes smaller. 

Algorithm~1, including these heuristics, is a minor adaption of a method suggested by \cite{flordal09} for incremental verification. They call it compositional verification, and the focus is on nonblocking and controllability properties, while the formulation here is adapted to incremental observer generation and specific observer properties based on transition systems. The main reason why this algorithm is presented here is that the nontrivial extension in \secr{mixed}, on observer abstraction for modular systems with shared unobservable events, can be computed in the same way. The difference is mainly that an additional observer operation is added on line 8.

\subsection{Transformation from forbidden state to nonblocking verification}\label{s:extobs}
The security related verification and synthesis problems considered in this paper are all related to identification of specific non-safe observer state properties, see Section~\ref{ssecprop}. In CSO, observer states that exclusively include secret states from the original system are non-safe, and in CSA, singleton observer states are considered as non-safe states. Non-safe states in an observer may formally be considered as {\em forbidden states}, and the verification as a forbidden state problem. This verification problem can be solved by introducing the state label $N$ for the non-safe states, and then use visible bisimulation as abstraction in Algorithm~1.

\fig{b}{    \begin{tikzpicture}[xscale=1, yscale=1, automaton]
            \node at (-3.3em,1.4em) {$G_i^d$};
            \node[initial, state, accepting] (0) at (0,0)    {$0$};
            \node[state] (00) at (7em,0em)    {$1$};
            \path[->]
              (0) edge [out=120,in=60, loop, distance=3em] node [above] {$\Sigma_i$} (0)
              (0) edge [] node [above] {$w_i$} (00);
           \end{tikzpicture}
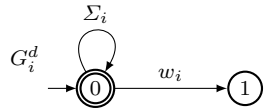
\vspace{0.25em}}{Detector automaton $G_i^d$ that inserts blocking states related to the forbidden states in $\mc O_{w_i}(G_i)$ by the extended observer $\mathcal O_e(G_i) = \mathcal O_{w_i}(G_i)\synch G_i^d$.}{ww}

\paragraph{Extended local observers} Since the problem only includes two types of states, safe and non-safe, an alternative to generic state labels and visible bisimulation is to transform the forbidden state problem to a {\em nonblocking problem}. All forbidden (non-safe) states in each individual observer $\mathcal O(G_i)$ are then augmented with a self-loop. For CSO these self-loops are labeled by $w_i$, $i = 1,\dots,n$, and the resulting local observers are called $\mathcal O_{w_i}(G_i)$.  For each such observer, a two-state {\em detector automaton} $G_i^d$, shown in \figr{ww}, is then introduced. It includes a marked state with a self-loop on the set of observable events $\Sigma_i$ in $G_i$ and a transition via the event $w_i$ to a non-marked state. The {\em extended local observer} 
\[
\mathcal O_e(G_i) = \mathcal O_{w_i}(G_i)\synch G_i^d
\]
then obtains non-marked blocking states added to every occurrence of a $w_i$ self-loop in $\mathcal O_{w_i}(G_i)$. Thus, every forbidden state in $\mathcal O(G_i)$  results in a direct transition to a blocking state in the extended local observer, while all original states in $\mathcal O_{w_i}(G_i)$ become marked in $\mathcal O_e(G_i)$. The reason is that no state in $\mathcal O_{w_i}(G_i)$ is explicitly marked, meaning that every state is implicitly considered to be marked in the synchronization. If any blocking states remain in the total extended  observer 
\beq
\mathcal O_e(G)=\mc O_e(G_1)\synch \mc O_e(G_2)\synch\cdots\synch \mc O_e(G_n),
\eeq{eq:obse}
this observer is blocking, and the observer $\mathcal O(G)$ includes one or more non-safe states from a CSO point of view. 

In the case of CSA, the transformation is simplified by choosing the same self-loop label $w$ for all observers $\mathcal O(G_i)$, $i = 1,\dots,n$, and the same $w$ label in every detector automaton $G_i^d$. This means that a blocking state may be reached first when all local observers have reached a non-safe state, a fact that is further motivated in \secr{subsec:anon_nonblock}. 

\paragraph{Abstraction preserving nonblocking} Conflict equivalence, introduced by \cite{Malik04}, preserves the nonblocking property of a transition system. This means that a system is nonblocking if and only if its conflict equivalence abstraction is also nonblocking. This abstraction, denoted $\mc A_c$, generally generates more efficient reductions compared to the visible bisimulation abstraction, here denoted $\mc A_v$. The reason is that only the nonblocking property is preserved by $\mc A_c$, while visible bisimulation, including divergence sensitivity, preserves temporal logics similar to \CTS\! \citep{bl_18}. 

By introducing the extended observer $\mc O_e$ as observer operator in Algorithm~1, an incremental observer based on the abstraction $\mc A_c$ is efficiently computed. This is possible, since conflict equivalence is congruent wrt hiding and synchronization \citep{Malik04}. The DES software tool Supremica \citep{olj:aff:sup:2006}  includes an incremental conflict equivalence implementation based on \citep{flordal09}. As an alternative, Algorithm~1 can also be implemented based on the visible bisimulation abstraction $\mc A_v$ and the original local observers including the non-safe state label $N$. Note that this abstraction is also congruent wrt hiding and synchronization \citep{bl:des:2019}.

The following example illustrates the transformation of a CSO verification problem to a nonblocking problem. Furthermore,  the efficiency of the conflict equivalence and the visible bisimulation abstractions is demonstrated.


\fig{t}{
         \begin{tikzpicture}[xscale=0.85, yscale=0.85, automaton]
            \node at (-2.5em,1.6em) {$G_i$};
            \node[state, initial] (0) at (0,0) {$0$};
            \node[state] (1) at (6.5em,0) {$1$};
            \node[state] (2) at (13em,0) {$2$};
            \node[state] (3) at (19.5em,0) {$3$};
            \path[->]
              (0) edge [out=30,in=150] node {$a_{i}$} (1)
              (1) edge [out=30,in=150] node {$a_{i}$} (2)
              (2) edge [out=30,in=150] node {$a_{i}$} (3)
              (3) edge [out=-150,in=-30] node [below] {$v_i$} (2)
              (2) edge [out=-150,in=-30] node [below] {$c_i$} (1)
              (1) edge [out=-150,in=-30] node [below] {$\xspa{0.7em}b_i,b_{i+1}$} (0);
          \end{tikzpicture}
\yspa{1.2em}

         \begin{tikzpicture}[xscale=0.85, yscale=0.85, automaton]
            \node at (-3.5em,1.6em) {$\mathcal O(G_i)$};
            \node[state, initial] (0) at (0,0) {$0$};
            \node[state] (1) at (6.5em,0) {$1$};
            \node[state, label=above: $\{N\}$] (2) at (13em,0) {$2$}; 
            \node[state, rounded rectangle] (3) at (20em,0) {$\{2,3\}$};
            \path[->]
              (0) edge [out=30,in=150] node {$a_{i}$} (1)
              (1) edge [out=30,in=150] node {$a_{i}$} (2)
              (1) edge [out=-150,in=-30] node [below] {$\xspa{0.7em}b_i,b_{i+1}$} (0)
              (2) edge node {$a_{i}$} (3)
              (2) edge [out=-150,in=-30] node [below] {$c_i$} (1)
              (3) edge [out=-120,in=-60, distance=3.7em] node [below] {$c_i$} (1)
              (3) edge [out=23,in=-23,loop,distance=2.45em] node [right] {$a_i$} (3);
          \end{tikzpicture}
\yspa{1.3em}

         \begin{tikzpicture}[xscale=0.85, yscale=0.85, automaton]
            \node at (-4.2em,1.6em) {$\mathcal O_{w_i}(G_i)$};
            \node[state, initial] (0) at (0,0) {$0$};
            \node[state] (1) at (6.5em,0) {$1$};
            \node[state] (2) at (13em,0) {$2$};
            \node[state, rounded rectangle] (3) at (20em,0) {$\{2,3\}$};
            \path[->]
              (0) edge [out=30,in=150] node {$a_{i}$} (1)
              (1) edge [out=30,in=150] node {$a_{i}$} (2)
              (1) edge [out=-150,in=-30] node [below] {$\xspa{0.7em}b_i,b_{i+1}$} (0)
              (2) edge node {$a_{i}$} (3)
              (2) edge [out=120,in=60,loop,distance=2.8em] node [above] {$w_i$} (2)
              (2) edge [out=-150,in=-30] node [below] {$c_i$} (1)
              (3) edge [out=-120,in=-60, distance=3.7em] node [below] {$c_i$} (1)
              (3) edge [out=23,in=-23,loop,distance=2.45em] node [right] {$a_i$} (3);
           \end{tikzpicture}
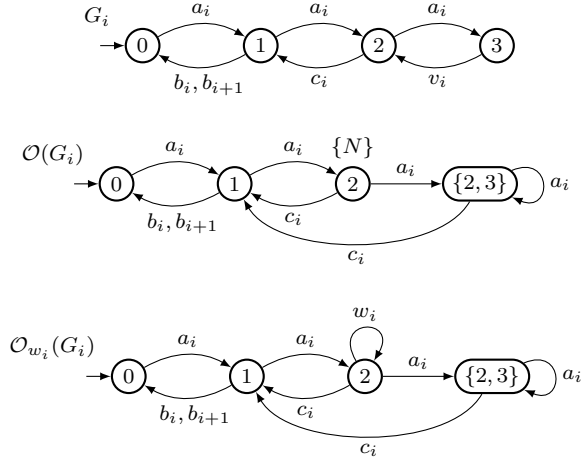
}{Transition system $G_i$ in Example~\ref{ex_obs_reduc}, its local observer $\mathcal O(G_i)$ including  state label $N$ at the non-safe state $2$, and local observer $\mathcal O_{w_i}(G_i)$ that is augmented with a $w_i$-self-loop at the non-safe state~$2$.}{fopi}

\fig{b}{
\xspa{0.2em}    
        \begin{tikzpicture}[xscale=1, yscale=1, automaton]
            \node at (-3.2em,1.4em) {$G_i^d$};
            \node[initial, state, accepting] (0) at (0,0)    {$0$};
            \node[state] (00) at (0em,-4.5em)    {$1$};
            \path[->]
              (0) edge [out=120,in=60, loop, distance=3em] node [above] {$a_i, b_i, b_{i+1}, c_i$} (0)
              (0) edge [] node [right] {$w_i$} (00);
           \end{tikzpicture}
\hspace{1em}
         \begin{tikzpicture}[xscale=0.85, yscale=0.85, automaton]
            \node at (-4.7em,1.6em) {$\mathcal O_e(G_i)$};
            \node[state, accepting, initial,rounded rectangle] (0) at (0,0) {$(0,0)$};
            \node[state,accepting,rounded rectangle] (1) at (6.5em,0) {$(1,0)$};
            \node[state,accepting,rounded rectangle] (2) at (13em,0) {$(2,0)$};
            \node[state, accepting,inner sep=1.4pt, rounded rectangle] (3) at (21em,0) {\fss $(\{2,\!3\},0)$};
            \node[state,rounded rectangle] (4) at (13em,5.5em) {$(2,1)$};
            \path[->]
              (0) edge [out=30,in=150] node {$a_{i}$} (1)
              (1) edge [out=30,in=150] node {$a_{i}$} (2)
              (1) edge [out=-150,in=-30] node [below] {$\xspa{0.7em}b_i,b_{i+1}$} (0)
              (2) edge node {$a_{i}$} (3)
              (2) edge [out=90,in=-90] node [right] {$w_i$} (4)
              (2) edge [out=-150,in=-30] node [below] {$c_i$} (1)
              (3) edge [out=-120,in=-60, distance=3.7em] node [below] {$c_i$} (1)
              (3) edge [out=20,in=-20,loop,distance=2.8em] node [right] {$a_i$} (3);
           \end{tikzpicture}}
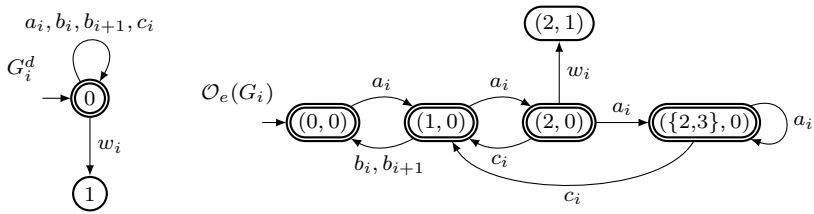
{The detector automaton $G_i^d$ that inserts an additional blocking state to the non-safe state $2$ in $\mathcal O_{w_i}(G_i)$, as depicted in $\mc O_e(G_i) = \mathcal O_{w_i}(G_i) \parallel G_i^d$.}{w_example}\vspace{1.7em}

\b{example}\label{ex_obs_reduc}
Consider the subsystem $G_i$, $i\in\mathbb{N}^+_n$ in \figr{fopi}, where $v_i$ is a local unobservable event and therefore replaced by $\veps$ before observer generation. The events $a_i$ and $c_i$ are local observable and the events $b_i$ and $b_{i+1}$ are observable but shared between neighbor subsystems, except the local events $b_1$ and $b_{n+1}$. The local observer $\mathcal O(G_i)$ is also shown in \figr{fopi}. 

The transition system $G_i$ is assumed to have one secret state, state $2$. Thus, the observer state $2$ is non-safe from a CSO point of view. This non-safe state with state label $N$ in $\mathcal O(G_i)$ is a forbidden state to which a $w_i$ self-loop is added in $\mathcal O_{w_i}(G_i)$  in \figr{fopi}. Including the detector automaton $G_i^d$ as depicted in \figr{w_example}, gives the extended local observer $\mathcal O_e(G_i)=\mathcal O_{w_i}(G_i) \synch G_i^d$ where the $w_i$ self-loop is replaced by a $w_i$ transition to a blocking state, also shown in \figr{w_example}.

In \tabr{topii_4_next} the complexity of the incremental extended observer $\mathcal O_e(G)^{\mathcal A_c}$, including abstraction  based on conflict equivalence,  is compared with the extended observer without abstraction $\mathcal O_e(G) \sa= \,\saa\parallel_{i\in\mathbb{N}^+_n}\smm \mathcal O_e(G_i)$ for different number of subsystems $n$. The result shows the strength of including the incremental abstraction, where the number of states $|X|$ and transitions $|T|$ including abstraction is constant independent of $n$, due to the specific structure of the problem.

Somewhat surprisingly, the incremental visible bisimulation abstraction $\mc A_v$ gives an even larger reduction down to only $2$ states and $2$ transitions, independent of $n$. This is shown in \citep{mona_18}. The reason why the conflict equivalence abstraction $\mc A_c$ does not achieve such an extreme reduction in this example is that the extended local observer $\mathcal O_e(G_i)$ includes an additional blocking state as a marker for the non-safe state. Thus, it is clear that for systems with special structures as the one in this example, visible bisimulation can be even more efficient than conflict equivalence.
\ebox
\e{example}


\tab{t}{Comparison of the number of states, transitions and the elapsed time after calculation of the abstracted extended observer $\mathcal O_e(G)^{\mathcal A_c}$ using incremental abstraction based on conflict equivalence, and the modular extended observer $\mathcal O_e(G)$ without abstraction.\hfill\yspa{2ex}}
{\begin{tabular}{c| ccc| lll}
    \toprule
  &  \multicolumn{3}{ c }{$\mathcal O_e(G)^{\mathcal A_c}$ \yspatb{0.85ex}{2ex}}   &\multicolumn{3}{ c }{$\mathcal O_e(G) \sa= \,\saa\parallel_{i\in\mathbb{N}^+_n}\smm \mathcal O_e(G_i)$} \\
\hline
     $n$  & $|\what X|$ & $|\what T|$ & $t_e$ (ms) & $|\what X|$ & $|\what T|$ & $t_e$ (s) \yspatb{0.6ex}{1ex} \\ \midrule 	
     3  & 8 & 17 & 3  & 125 & 585 & 0.028   \\
     5  & 8 & 17 & 5  & 3,125 & 23,625 & 0.082\\
     8  & 8 & 17 & 9  & 390,625 & 4,640,625 & 5.49 \\
     9  & 8 & 17 & 17 & 1,953,125 & 26,015,625 & 24.32 \\
     12 & 8 & 17 & 18  & $\approx 50 \times 10^6$ & -- & --\\ \bottomrule
\end{tabular}}{topii_4_next}

\section{Opacity and anonymity for modular systems} \label{op_an}
So far we have shown how a specific type of states in an observer called non-safe states can be identified in an efficient way for modular systems. In this section a more detailed definition of these non-safe states is given. This is done for the security and privacy problems current state opacity and anonymity, focusing on modular structures. 

A centralized architecture is considered, including one single intruder of the system. It is assumed that the intruder has full knowledge of the system structure. However, it can only observe a subset of the system events, included in the set of observable events. Based on its observations, the intruder is assumed to be able to construct an observer of the system, where only observable events are included as transition labels.

\subsection{Current state opacity and anonymity}
In current-state opacity (CSO) \citep{Saboori:2014,lesage16},  the goal is to evaluate if it is possible to estimate any secret states in a system based on its observable events. For a transition system $G$, let $X_S\subseteq X$ be the set of secret states. 
This system is then said to be {\em opaque} if for every string of observable events $s\in \mc L(G)$, each corresponding state set $Y=\delta(I,s)$ that includes secret states also includes at least one non-secret state from the set $X\smm\setminus \smm X_S$. Note that the natural projection on observable events that often is used in opacity definitions, see for instance  \cite{lesage16}, is not involved in this section, since the strings $s\in\mc L(G)$ only include observable events. The unobservable events are replaced by $\veps$.

\begin{definition}[Current state opacity] \label{d:cso}  \xspa{0.1em}
{\rm
Consider a nondeterministic transition system~$G$, where any unobservable events are replaced by $\veps$ and $X_S\subseteq X$ is the set of secret states. For a string of observable events $s\in \mc L(G)$, the block state $Y=\delta(I,s)$ is {\em safe} if $Y\nsubseteq X_S$, and $G$ is {\em current state opaque} if for all strings $s\in \mc L(G)$, all corresponding block states $Y=\delta(I,s)$ are safe.
\ebox}
\end{definition}
Since the block states $Y=\delta(I,s)$ in this CSO definition are states in the corresponding observer $\mc O(G)$, the following proposition follows immediately.
\begin{proposition}[Current state opacity and safe/non-safe observer states] \label{p:cso}  \xspa{0.1em}
{\rm
A transition system $G$ is \emph{current-state opaque} if and only if all  states in the observer $\mathcal O(G)$ are safe. Furthermore, a  state $Y$ in $\mathcal O(G)$ is {\em non-safe} if and only if it includes only secret states from $G$, \ie $Y\subseteq X_S$.
\ebox}
\end{proposition}
According to this proposition, a transition system $G$ is current-state non-opaque if and only if at least one state in the observer $\mathcal O(G)$  includes only secret states from $G$, and is therefore {\em non-safe}. The label $N$ is a state label for all non-safe states in  $\mc O(G)$.

\paragraph{Current state anonymity} With the increasing popularity of location-based services for mobile devices, privacy concerns about the unwanted revelation of user's current location are raised. For this reason the notion of CSO is adapted, and a new related notion called \emph{current state anonymity} (CSA) is introduced \citep{Wu_2014}. CSA captures the observer's inability to know for sure the current locations of  moving patterns.

\begin{definition}[Current state anonymity] \label{d:csa}  \xspa{0.1em}
{\rm
Consider a nondeterministic transition system~$G$, where any unobservable events are replaced by $\veps$. For a string of observable events $s\in \mc L(G)$, the block state $Y=\delta(I,s)$ is {\em safe} if this state set is not a singleton ($|Y|>1$). Furthermore, $G$ is {\em current state anonymous} if for all strings $s\in \mc L(G)$, all corresponding block states $Y=\delta(I,s)$ are safe.
\ebox}
\end{definition}
In the same way as for CSO, the block states $Y=\delta(I,s)$ in this CSA definition are states in the corresponding observer $\mc O(G)$, which directly implies the following proposition.
\begin{proposition}[Current state anonymity and safe/non-safe observer states] \label{p:csa}  \xspa{0.1em}
{\rm
A transition system $G$ is \emph{current-state anonymous} if and only if all block states in the observer $\mathcal O(G)$ are safe. Furthermore, a block state $Y$ in $\mathcal O(G)$ is {\em safe} if and only if it is not a singleton ($|Y|>1$).
\ebox}
\end{proposition}
According to this proposition, a transition system $G$ is current-state non-anonymous, if and only if at least one block state $Y$ in the observer $\mathcal O(G)$ is a singleton and is therefore {\em non-safe}. Obviously, anonymity is evaluated by verifying that no observer block state is a  singleton state. This is natural, since more than one system state in each observer block state implies an uncertainty in determining the exact location of a moving pattern. Finally, in the same way as for CSO,  the label $N$ is a state label for all non-safe (non-anonymous) block states in $G$ and corresponding states in $\mc O(G)$.

The following example shows the observers for CSO and CSA, including their different  $N$ (non-safe) state label interpretations. 

\b{example}\label{ex_abs_oa}
Consider the transition system $G$ in \figr{oa_obs} where the secret state set $X_S=\{0,1,2\}$. The event $u$ is unobservable and is therefore replaced by $\veps$ before the observer generation, where the source and target states of the~$\veps$ transition are merged. Although the observers are structurally equal, depending on the verification problem, the interpretation differs concerning the non-safe states and therefore the state labeling. The block state $\{1,2\}$ in the observer $\mathcal O_{\tiny\mbox{CSO}}(G)$  has label $N$, because both states $1$ and $2$ are secret states. On the other hand, the corresponding state in $\mathcal O_{\tiny\mbox{CSA}}(G)$ does not have state label~$N$, as it is not a singleton state. 

Since $G$ does not include any subsystems, all events can be considered as local, since no synchronization between local subsystems is performed.  The observable events $a$ and $b$ are therefore hidden by relabeling them with $\tau$. In the visible bisimulation (VB) abstraction $\mathcal O_{\tiny\mbox{CSO}}(G)^{\mathcal A^{\{a,b\}}}$\!, the $N$-labeled states with a $\tau$ transition in between are merged, while no reduction is achieved for  $\mathcal O_{\tiny\mbox{CSA}}(G)^{\mathcal A^{\{a,b\}}}$\!. We notice that only states with the same state label (label $N$ or no label) are merged in the VB abstraction.
\hfill \ebox\e{example}

\fig{t}{   
\xspa{0.1em}
\begin{tikzpicture}[->,xscale=1, yscale=1, automaton]
            \node at (-2.5em,1.3em) {$G$};
            \node[initial, state] (1) at (0,0) {$0$};
            \node[state] (2) at (5em,0em)    {$1$};
            \node[state] (3) at (10em,0em)    {$2$};
            \node[state] (4) at (15em,0em)    {$3$};
            \node at (21.5em,0em)    {$X_S=\{0,1,2\}$};
            \path[->]
              (1) edge node [above] {$a$} (2)
              (2) edge node [above] {$u$} (3)
              (3) edge node [above] {$b$} (4);
\end{tikzpicture}
\yspa{1.6em}

\begin{tikzpicture}[xscale=1, yscale=1, automaton]
            \node at (-3.5em,1.3em) {$\mathcal O_{\tiny\mbox{CSO}}(G)$};
            \node[initial, state,label=$\{N\}$,rounded rectangle] (1) at (0,0) {$\{0\}$};
            \node[state,label=$\{N\}$,rounded rectangle] (2) at (6em,0em)    {$\{1,2\}$};
            \node[state,rounded rectangle] (3) at (12em,0em)    {$\{3\}$};
            \path[->]
              (1) edge node [above] {$a$} (2)
              (2) edge node [above] {$b$} (3);
\end{tikzpicture}
\hspace{1.7em}
\begin{tikzpicture}[xscale=1, yscale=1, automaton]
\node at (-3.5em,1.3em) {$\mathcal O_{\tiny\mbox{CSA}}(G)$};
            \node[initial, state,label=$\{N\}$,rounded rectangle] (1) at (0,0) {$\{0\}$};
            \node[state,rounded rectangle] (2) at (6em,0em)    {$\{1,2\}$};
            \node[state,label=$\{N\}$, rounded rectangle] (3) at (12em,0em)    {$\{3\}$};
            \path[->]
              (1) edge node [above] {$a$} (2)
              (2) edge node [above] {$b$} (3);
\end{tikzpicture}
\yspa{1.8em}

\begin{tikzpicture}[xscale=1, yscale=1, automaton]
            \node at (-5.75em,1.3em) {$\mathcal O_{\tiny\mbox{CSO}}(G)^{\mathcal A^{\{a,b\}}}$};
            \node[initial, state,label=$\{N\}$,rounded rectangle] (1) at (0,0) {$\{0,1,2\}$};
            \node[state,rounded rectangle] (2) at (6.5em,0em)    {$\{3\}$};
            \path[->]
              (1) edge node [above] {$\tau$} (2);
\end{tikzpicture}
\hspace{1.5em}
\begin{tikzpicture}[xscale=1, yscale=1, automaton]
\node at (-4.8em,1.3em) {$\mathcal O_{\tiny\mbox{CSA}}(G)^{\mathcal A^{\{a,b\}}}$};
            \node[initial, state,label=$\{N\}$,rounded rectangle] (1) at (0,0) {$\{0\}$};
            \node[state,rounded rectangle] (2) at (6em,0em)    {$\{1,2\}$};
            \node[state,label=$\{N\}$, rounded rectangle] (3) at (12em,0em)    {$\{3\}$};
            \path[->]
              (1) edge node [above] {$\tau$} (2)
              (2) edge node [above] {$\tau$} (3);
\end{tikzpicture}
\yspa{0.2em}}
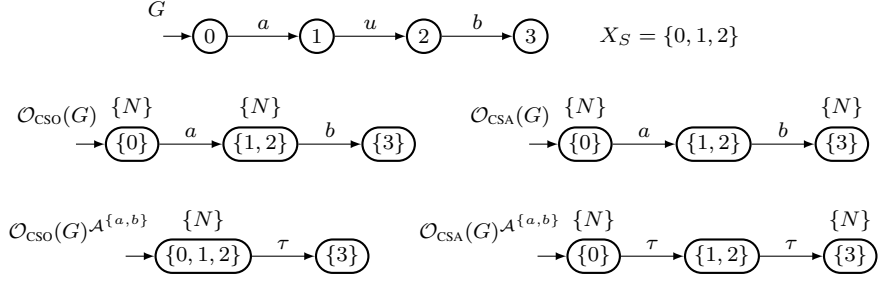
{A system model $G$, its two observers $\mathcal O_{\tiny\mbox{CSO}}(G)$ and $\mathcal O_{\tiny\mbox{CSA}}(G)$ when the event $u$ is unobservable, and corresponding VB abstractions $\mathcal O_{\tiny\mbox{CSO}}(G)^{\mathcal A^{\{a,b\}}}$ and $\mathcal O_{\tiny\mbox{CSA}}(G)^{\mathcal A^{\{a,b\}}}$ when $a$ and $b$ are local observable events.}{oa_obs} 

To summarize this subsection,  a block state $Y$ in the observer $\mathcal O(G)$ is  \emph{non-safe} and is augmented with state label $N$, in CSO verification when $Y\subseteq  X_S$, and in CSA verification when $|Y|=1$. These results are now generalized to modular systems.

\subsection{Current state opacity and anonymity for modular systems}
For a modular transition system $G\sa= \,\saa\parallel_{i\in\mathbb{N}^+_n}\smm G_i$, CSO requires a modified definition of secret states. No shared unobservable events also means that both CSO and CSA can be expressed in terms of safety of the local block states $Y_i$ in $G_i$. 

Before CSO is defined for modular systems, consider an $n$-dimensional cross product $Y=Y_1\times Y_2 \times \cdots \times Y_n$, where the $i$-th set $Y_i$ is replaced by the set $Z_i$. This modified cross product is denoted 
\beq
R(Y,Z_i)=Y_1\times \cdots \times Y_{i-1} \times Z_i \times Y_{i+1} \times \cdots \times Y_n
\eeq{eq:RYZ}

\begin{definition}[Current state opacity for modular systems] \label{d:csoo}  
{\rm
Consider a modular transition system $G\sa= \,\saa\parallel_{i\in\mathbb{N}^+_n}\smm G_i$, where $X_{S_i}$ is the set of secret states for subsystem $G_i$. The set of secret states for $G$ with state space $X=X_1\times\cdots\times X_n$ is then defined as
\[
X_S=\bigcup_{i=1}^n R(X,X_{S_i})
\]
For a string of observable events $s\in \mc L(G)$, the block state $Y=\delta(I,s)$ of the modular system $G$ is {\em safe} if  $Y\nsubseteq X_S$, and $G$ is {\em current state opaque} if for all strings $s\in \mc L(G)$, all corresponding block states $Y=\delta(I,s)$ are safe.
\ebox}
\end{definition}
The only difference between this CSO definition and \defr{d:cso} is the more complex structure of the set of secret states $X_S$, which is illustrated in the following example.

\fig{b}{   
\xspa{2em}
\begin{tikzpicture}[->,xscale=1, yscale=1, automaton]
            \node at (-2.7em,1.3em) {$G_1$};
            \node[initial, state] (0) at (0,0) {$0$};
            \node[state] (1) at (5em,0em)    {$1$};
            \node[state] (2) at (10em,0em)    {$2$};
            \node[state] (3) at (15em,0em)    {$3$};
            \node at (21.3em,0em)    {$X_{S_1}=\{1,2\}$};
            \path[->]
              (0) edge node [above] {$a$} (1)
              (1) edge node [above] {$b$} (2)
              (2) edge node [above] {$v$} (3);
\end{tikzpicture}
\yspa{1.3em}

\xspa{2em}
\begin{tikzpicture}[->,xscale=1, yscale=1, automaton]
            \node at (-2.7em,1.3em) {$G_2$};
            \node[initial, state] (0) at (0,0) {$0$};
            \node[state] (1) at (5em,0em)    {$1$};
            \node[state] (2) at (10em,0em)    {$2$};
            \node[state] (3) at (15em,0em)    {$3$};
            \node[state] (4) at (20em,0em)    {$3$};
            \node at (26em,0em)    {$X_{S_2}=\{4\}$};
            \path[->]
              (0) edge node [above] {$a$} (1)
              (1) edge node [above] {$u$} (2)
              (2) edge node [above] {$b$} (3)
              (3) edge node [above] {$v$} (4);
\end{tikzpicture}
\yspa{1.7em}

\begin{tikzpicture}[xscale=1, yscale=1, automaton]
            \node at (-4.7em,1.3em) {$\mc O(G_1\synch G_2)$};
            \node[initial, state,rounded rectangle] (0) at (0,0) {$Y$};
            \node[state,label=$\{N\}$,rounded rectangle] (1) at (5.05em,0em)    {$Y'$};
            \node[state,label=$\{N\}$,rounded rectangle] (2) at (10.1em,0em)    {$Y''$};
            \path[->]
              (0) edge node [above] {$a$} (1)
              (1) edge node [above] {$b$} (2);
\end{tikzpicture}
\yspa{0.4em}\xspa{4.8em}}
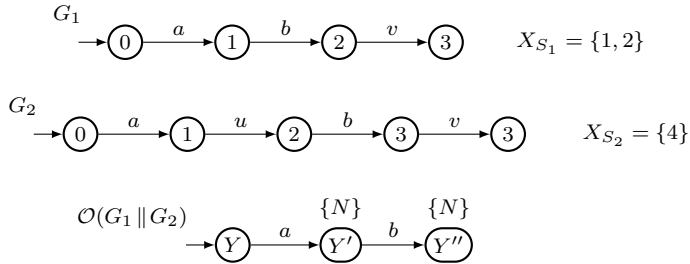
{Two subsystems $G_1$ and $G_2$ with observable events $a$ and $b$, and unobservable events $u$ and $v$, as well as the observer of the composed system $\mc O(G_1\synch G_2)$ where $Y=\{(0,0)\}$, $Y'=\{(1,1),(1,2)\}$, and $Y''=\{(2,3),(3,4)\}$.}{cso_mod}

\b{example}\label{e:cso_mod}
Consider the subsystems $G_1$ and $G_2$ in \figr{cso_mod}, where the events $a$ and $b$ are observable, and $u$ and $v$ are unobservable. Since $v$ is shared,  local observers can not be generated. Thus, $G_1$ and $G_2$ are first synchronized before the unobservable events are replaced by $\veps$, and the observer $\mc O(G_1\synch G_2)$ is generated where the three block states are $Y=\{(0,0)\}$, $Y'=\{(1,1),(1,2)\}$, and $Y''=\{(2,3),(3,4)\}$. The local event $u$ in $G_2$ gives the two states in $Y'$, and the shared event $v$ gives the two states in $Y''$.

The set of secret states $X_S=X_{S_1}\times X_{2} \cup X_{1}\times X_{S_2}=\{1,2\}\times\{0,1,2,3,4\}\cup \{0,1,2,3,4\}\times \{4\}$ implies that $Y'\subseteq X_S$ and $Y''\subseteq X_S$, while $Y\nsubseteq X_S$. Thus, $Y'$ and $Y''$ are non-safe states, which means that the $G_1\synch G_2$ is current state non-opaque. Due to the union in the definition of secret states in \defr{d:csoo}, it is enough that $G_1$ is in the local secret state $1$ to make $Y'$ a non-safe state in the composed system. The state~$Y''$ is non-safe due to a more complex behavior, where first $G_1$ is in the local secret state $2$ and $G_2$ is in the non-secret state $3$. After the shared unobservable event~$v$ has been executed, the opposite occurs where $G_2$ is instead in a local secret state (state $4$), and~$G_1$ is in a non-secret state (state $3$). 

Both in state $Y'$ and $Y''$ an intruder is able to detect that one of the subsystems is in a secret state. In the non-safe state $Y'$, the intruder knows that $G_1$ is in a secret state. A safe state implies that both a secret and a non-secret state can be occupied. Thus, there are one or more states in a safe block state where no subsystem is in a secret state. In $Y''$ one of the subsystems is in a secret state, but the shared unobservable event $v$ does not make it possible to determine if it is $G_1$ or $G_2$, only that one of them is in its secret state.
\ebox
\e{example}

In the next example, it is also shown how the CSO definition in \defr{d:csoo} can be simplified for modular systems where all unobservable events are local.

\fig{t}{   
\begin{tikzpicture}[xscale=1, yscale=1, automaton]
            \node at (-3.9em,1.3em) {$\mc O(G_1)$};
            \node[initial, state,rounded rectangle] (0) at (0,0) {$\{0\}$};
            \node[state,label=$\{N\}$,rounded rectangle] (1) at (5.45em,0em)    {$\{1\}$};
            \node[state,rounded rectangle] (2) at (11.4em,0em)    {$\{2,3\}$};
            \path[->]
              (0) edge node [above] {$a$} (1)
              (1) edge node [above] {$b$} (2);
\end{tikzpicture}
\hspace{1.2em}
\begin{tikzpicture}[xscale=1, yscale=1, automaton]
            \node at (-3.9em,1.3em) {$\mc O(G_2)$};
            \node[initial, state,rounded rectangle] (0) at (0,0) {$\{0\}$};
            \node[state,rounded rectangle] (1) at (5.92em,0em)    {$\{1,2\}$};
            \node[state,rounded rectangle] (2) at (12.4em,0em)    {$\{3,4\}$};
            \path[->]
              (0) edge node [above] {$a$} (1)
              (1) edge node [above] {$b$} (2);
\end{tikzpicture}
\yspa{0.4em}}
{Local observers for the subsystems $G_1$ and $G_2$ in \figr{cso_mod} when the shared unobservable event $v$ in $G_2$ is replaced by the local unobservable event $w$.}{cso_modd}

\b{example}\label{e:cso_modd}
Consider the subsystems $G_1$ and $G_2$ in \figr{cso_mod}, where the shared unobservable event $v$ in $G_2$ is replaced by the local unobservable event $w$. No shared unobservable events give the local observers in \figr{cso_modd}. Synchronization of these observers results in the three states $(Y_1,Y_2)\define (\{0\},\{0\})$, $(Y'_1,Y'_2)\define (\{1\},\{1,2\})$, and $(Y''_1,Y''_2)\define (\{2,3\},\{3,4\})$ in $\mc O(G_1)\synch \mc O(G_2)$. 

Taking the union of the state labels in the synchronization according to \defr{dsynch}, the second state $(Y'_1,Y'_2)$ becomes non-safe since the second state $Y'_1=\{1\}$ in $G_1$ is non-safe, while the rest of the states are safe. This result coincides with \defr{d:csoo}, which in the same way as in \exr{e:cso_mod} shows that replacing the shared unobservable event~$v$ with the local event $w$ in $G_2$ means that only the second state $Y'$ in $\mc O(G_1\synch G_2)$ is non-safe.
\ebox
\e{example}

This example illustrates that the existence of non-safe block states for modular systems with only local unobservable events can be decided based on the existence of local non-safe block states. The following proposition confirms this statement.

\begin{proposition}[Non-safe block states for modular systems] \label{p:cso_mod}  \xspa{0.1em}
{\rm
Consider a modular transition system $G\sa= \,\saa\parallel_{i\in\mathbb{N}^+_n}\smm G_i$, where $X_{S_i}$ is the set of secret states for subsystem~$G_i$, and all unobservable events are local.  For a string of observable events $s\in \mc L(G)$, the block state $Y=Y_1\times \cdots \times Y_n=\delta(I,s)$ of the modular system~$G$ is {\em non-safe}, that is $Y\subseteq X_S=\bigcup_{i=1}^n R(X,X_{S_i})$, if and only if $(\exists i\in \mathbb{N}^+_n) \,Y_i\subseteq X_{S_i}$.}

{Proof:} {\rm  ($\IMP$) Assume by contradiction that $(\forall i\in \mathbb{N}^+_n) \,Y_i\nsubseteq X_{S_i}$, which means that $(\forall i\in \mathbb{N}^+_n) \,Y_i\cap X_i\setm X_{S_i} \neq \varnothing$. By the notation \rf{eq:RYZ}, this assumption yields $Y\cap X\setminus X_S$ $=Y\cap X\setminus \bigcup_{i=1}^n R(X,X_{S_i}) =Y \cap\, \bigcap_{i=1}^n R(X,X_i\setminus X_{S_i}) =\bigcap_{i=1}^n R(Y,Y_i\cap X_i\setminus X_{S_i}) \neq \varnothing$, and we conclude that $Y\nsubseteq X_S$, which is a contradiction.

($\Leftarrow$) Since $(\forall j\in \mathbb{N}^+_n) \,Y_j\subseteq X_j$, it follows that $Y=R(Y,Y_i)\subseteq R(X,Y_i)$. Furthermore, assuming that $(\exists i\in \mathbb{N}^+_n) \,Y_i\subseteq X_{S_i}$, and selecting $i$ such that $Y_i\subseteq X_{S_i}$, it implies that $Y\subseteq R(X,Y_i) \subseteq R(X,X_{S_i}) \subseteq \bigcup_{i=1}^n R(X,X_{S_i}) = X_S$.
\ebox}
\end{proposition}

A block state $Y=Y_1\times \cdots \times Y_n$ in $G=\parallel_{i\in\mathbb{N}^+_n}\smm G_i$ is also a state in the observer $\mc O(\parallel_{i\in\mathbb{N}^+_n}\smm G_i)$, and according to \lemr{tobs} the block state $Y_i$  is then also a state in the corresponding local observer $\mc O(G_i)$. Thus, a modular system $G$ without shared unobservable events is CSO if there is no non-safe state in any local observer. The system is still opaque if the states in the global observer that include non-safe components are not reachable.

The interpretation of CSO from an intruder perspective was discussed in \exr{e:cso_mod}. For a system to be in a non-safe state $Y$, it is obviously enough, according to this example and \defr{d:csoo}, that one of the subsystems is in a secret state for every state $x\in Y$. Consider, for instance, that each subsystem models one moving pattern, say a person. It is then not necessary that all persons are in local secret states to get a global non-safe state. It can be enough with one person, depending on the availability of observable events. In another scenario, where only one person is involved, the person being in a secret state in one subsystem means that the  local states of the other subsystems then represent the absence of the single person. This modeling scenario is applied in the multiple floor/elevator building in \secr{sec:ex}. 

Generally, the modular system model in \defr{d:csoo} is very flexible. The definition of secret states is often natural, as we have tried to motivate above, but alternative definitions are possible, for instance by replacing the union operator with the intersection operator, such that $X_S=\bigcap_{i=1}^n R(X,X_{S_i})=X_{S_1}\times \cdots \times X_{S_n}$. A similar type of intersection is applied when current state anonymity is analyzed for modular systems.  

\paragraph{Current state anonymity for modular systems} In the following definition, all subsystems must be in a singleton state for the whole system to break the location privacy for moving patterns.

\begin{definition}[Current state anonymity for modular systems] \label{d:csaa}  
{\rm
Consider a modular system $G\sa= \,\saa\parallel_{i\in\mathbb{N}^+_n}\smm G_i$. For a string of observable events $s\in \mc L(G)$, the global block state $Y=\delta(I,s)$ is {\em non-safe} if $Y=x\in X$, \ie $Y$ is a tuple of singleton states. Furthermore, $G$ is {\em current state anonymous} if for all strings $s\in \mc L(G)$, all corresponding global block states $Y=\delta(I,s)$ are safe, \ie no global state is a tuple of singleton states.
\ebox}
\end{definition}
According to this definition, a global block state is only non-safe when all local states are singleton states. The motivation for this interpretation is that the synchronized subsystems together are assumed to model a map.  Location privacy is then violated if it is possible to get a specific location on such a map. This corresponds to a global singleton state, involving singleton states for all subsystems.

For a string of observable events $s\in \mc L(G)$, the global block state $Y=\delta(I,s)$ in $G\sa= \,\saa\parallel_{i\in\mathbb{N}^+_n}\smm G_i$ is also a state in the observer $\mc O(G)$. Furthermore, for a modular system without shared unobservable events,  \lemr{tobs} implies that $\mc O(G)=\parallel_{i\in\mathbb{N}^+_n}\smm \mc O(G_i)$. The state $Y$ can then also be expressed as $Y=Y_1\times \cdots \times Y_n$, where $Y_i$  is a state in the corresponding local observer $\mc O(G_i)$. 

One clear difference between CSO and CSA is, however, the handling of state labels in the synchronization of the local observers $\parallel_{i\in\mathbb{N}^+_n}\smm \mc O(G_i)$. In the case of CSA, it is then necessary to take the {\em intersection} of the actual non-safe state labels $N$ from the individual subsystems, to generate a correct global state label according to \defr{d:csaa}. Thus, in current state anonymity for modular systems, the union of non-safe state labels $N$ in the synchronous composition of observers is replaced by the intersection of these state labels from the individual subsystems. The following example illustrates the differences between CSA and CSO.

\b{example}
First, consider the observer $\mc O(G)$ in \figr{cso_mod}. The structure of this observer is the same as for CSA, but the non-safe states are different. Concerning CSA, the first state $Y$ is non-safe, since it is a singleton state, while the states $Y'$ and $Y''$ are safe. 

In \figr{csa_mod} no shared unobservable event is involved, and therefore the observer can be generated as $\mc O(G_1)\synch \mc O(G_2)$. In the local observers the non-safe states are assigned to the singleton states. After synchronization, taking the intersection of the~$N$ state labels, only the first state in $\mc O(G_1)\synch \mc O(G_2)$ is non-safe. This is confirmed by noting that it is also the only singleton state in $\mc O(G_1)\synch \mc O(G_2)$.
\ebox
\e{example}

\fig{t}{   
\begin{tikzpicture}[xscale=1, yscale=1, automaton]
            \node at (-3.9em,1.3em) {$\mc O(G_1)$};
            \node[initial, state, label=$\{N\}$, rounded rectangle] (0) at (0,0) {$\{0\}$};
            \node[state, label=$\{N\}$, rounded rectangle] (1) at (5.45em,0em)    {$\{1\}$};
            \node[state, rounded rectangle] (2) at (11.4em,0em)    {$\{2,3\}$};
            \path[->]
              (0) edge node [above] {$a$} (1)
              (1) edge node [above] {$b$} (2);
\end{tikzpicture}
\hspace{1.2em}
\begin{tikzpicture}[xscale=1, yscale=1, automaton]
            \node at (-3.9em,1.3em) {$\mc O(G_2)$};
            \node[initial, label=$\{N\}$, state,rounded rectangle] (0) at (0,0) {$\{0\}$};
            \node[state, rounded rectangle] (1) at (5.92em,0em)    {$\{1,2\}$};
            \node[state, rounded rectangle] (2) at (12.4em,0em)    {$\{3,4\}$};
            \path[->]
              (0) edge node [above] {$a$} (1)
              (1) edge node [above] {$b$} (2);
\end{tikzpicture}
\yspa{0.4em}}
{Local observers for the subsystems $G_1$ and $G_2$ in \figr{cso_mod} when the shared unobservable event $v$ in $G_2$ is replaced by the local unobservable event $w$. The non-safe state labels are determined based on CSA.}{csa_mod}

\subsection{Transformation of current state opacity and anonymity to nonblocking problems} \label{subsec:anon_nonblock}
In \secr{s:extobs} it was shown how CSO and CSA verification can be transformed to nonblocking problems, when all unobservable events are local. More specifically, a system $G$ is CSO if the observer $\mathcal O_e(G)$ in \rf{eq:obse} is nonblocking. For CSA, the same self-loop label $w$ in all observers $\mathcal O_e(G_i)$, $i = 1,\dots,n$ means that a blocking state is only reached when all local observers have reached a non-safe singleton state. This models the intersection of all non-safe state labels $N$ in the synchronization of the individual subsystems, that is required to reach a global non-safe state in the case of~CSA.

\subsection{Other types of opacity} \label{subsec:LBO}
Opacity can also be defined based on languages, see \cite{dubreil08}, \cite{Badouel:2007}, and \cite{Lin_2011}. For a system $G$ with a set of initial states $I$ and a language $\mc L(G,I)$, two sublanguages are introduced, a secret language $L_S\subseteq \mc L(G,I)$ and a non-secret language $L_{NS} \subseteq \mc L(G,I)$, where $L_{S}\cap L_{NS}=\varnothing$.  Unobservable events are here not replaced by $\veps$. Instead, a projection~$P$ from all events to the observable events is introduced. The system $G$ is then \emph{language-based opaque} if $L_S\subseteq P^{-1}[P(L_{NS})]$.  

To verify language-based opacity (LBO), this formulation can be transformed to CSO as in \citep{wu2013}, and then verified based on the techniques proposed in this paper. This includes a modular formulation of the transformation from LBO to CSO. Furthermore, two notions of initial state opacity (ISO) and initial/final state opacity (IFO), as presented in \citep{wu2013}, can also be transformed to LBO and then to a CSO problem.

\section{Observer abstraction for systems with shared unobservable events} \label{mixed}
For modular systems with partial observation and shared unobservable events, the observer generation including incremental abstraction presented in \secr{obs:gen} and Algorithm 1 must be reformulated. The problem is that the complete observer can not be computed by only synchronizing the local observers as in \rf{obs_eq}. The reason is that shared unobservable events can not be replaced by $\veps$ before they have become local after synchronization.

\subsection{Incremental observer generation}
To highlight this complication, the more expressive observer operator $\mc O_{\Sigma^\veps}(G)$ is used, where we remind that the subscript $\Sigma^{\veps}$ includes the set of local unobservable events that are replaced by $\veps$ before the observer generation. First, the following lemma shows that an observer can also be generated incrementally. An observer $\mc O_{\Sigma^\veps_1}(G)$ is then computed, assuming that the events in $\Sigma^\veps_1$ are local and unobservable. When additional local events in $\Sigma^{\veps}_{2}$ are considered, an update of this observer as $\mc O_{\Sigma^\veps_2}(\mc O_{\Sigma^\veps_1}(G))$ is shown to give the same result as generating the observer $\mc O_{\Sigma^\veps_1 \dot\cup \Sigma^\veps_2}(G)$ in one step.

\begin{lemma}[Incremental observer generation]\label{lemma:obs}
{\rm Consider a nondeterministic transition system $G$ with a state set $X$, an initial state set $I$, a set of observable events $\Sigma^o$, and a set of unobservable  events $\Sigma^{uo}$. Let $\Sigma^\veps_1\subseteq \Sigma^{uo}$, $\Sigma^\veps_2\subseteq \Sigma^{uo}$, and $\Sigma^\veps= \Sigma^\veps_1\,\dot\cup \,\Sigma^\veps_2$ be sets of unobservable events that are replaced by $\veps$ before  corresponding observer generation. Then }
\[
\mc O_{\Sigma^{\veps}}(G) = \mc O_{\Sigma^{\veps}_2}(\mc O_{\Sigma^{\veps}_1}(G)).
\]

{Proof:} {\rm  For the total event set $\Sigma=\Sigma^o\cup\Sigma^{uo}$, consider the language $\mc L(G) \subseteq \Sigma^*$ and the projections $P_1:\Sigma^* \ra (\Sigma \setminus\Sigma^\veps_1)^*$, $P_2:(\Sigma \setminus \!\Sigma^\veps_1)^* \ra (\Sigma \setminus\!(\Sigma^\veps_1\dot\cup \Sigma^\veps_2))^*$, and  \lb$P:\Sigma^* \ra (\Sigma \setminus\!(\Sigma^\veps_1\dot\cup \Sigma^\veps_2))^*$. After a string $s\in\mc L(G)$ has been executed, the block state in the observer $\mc O_{\Sigma^\veps_1} (G)$ can be expressed as $Y_{1} = \{x\in X\st (\exists x_0\in I) $\lb$x_0\stackrel{P_{1}(s)}{\Longrightarrow} x\}$. The corresponding string $t=P_1(s)$, executed by the observer \lb $\mc O_{\Sigma^{\veps}_2}(\mc O_{\Sigma^{\veps}_1}(G))$, results in the block state $Y_{2} = \{Y_{1}\in 2^X\st \what I_{1}\stackrel{P_2(t)\yspa{0.1ex}}{\Longrightarrow} Y_{1} \}$, where $\what I_1$ is the initial state of $\mc O_{\Sigma^\veps_1} (G)$. Moreover, the projection $P(s)$  generates the block state $Y = \{x\in X\st (\exists x_0\in I)\,x_0\stackrel{P(s)}{\Longrightarrow} x\}$ in the observer $\mc O_{\Sigma^\veps} (G)$. The bijective function $f:2^{2^X}\ra 2^X$, where $f(Y_{2})=\bigcup_{Y_{1}\in Y_{2}} Y_{1}$, together with the fact that $P_2(t)=P_2(P_1(s))=P(s)$, finally means that $Y=f(Y_2)$ for any string $s\in\mc L(G)$. Hence, the states in the observers $\mc O_{\Sigma^{\veps}_2}(\mc O_{\Sigma^{\veps}_1}(G))$ and $\mc O_{\Sigma^{\veps}}(G)$ are isomorphic, and the observers are therefore structurally equal.

The equality also includes the state labels. First consider the default assumption on union of state labels in observer block states, and the block state relation $Y=f(Y_2)=\bigcup_{Y_1\in Y_2} Y_1$. Then the state label of a block state $Y_2$ in $\mc O_{\Sigma^{\veps}_2}(\mc O_{\Sigma^{\veps}_1}(G))$ is 
$
\bigcup_{Y_1\in Y_2} \bigcup_{x\in Y_1} \lambda(x)=\bigcup_{x\in\left(\bigcup_{Y_1\in Y_2} Y_1\right)} \lambda(x)= \bigcup_{x\in Y} \lambda(x), 
$ 
which is the state label of the corresponding block state $Y=f(Y_2)$ in $\mc O_{\Sigma^{\veps}}(G)$. Thus, the state labels for the two observers coincide. The alternative interpretation for CSO, where union is replaced by intersection in the observer generation, gives the same result, since then 
$
\bigcap_{Y_1\in Y_2} \bigcap_{x\in Y_1} \lambda(x)=\bigcap_{x\in\left(\bigcup_{Y_1\in Y_2} Y_1\right)} \lambda(x)= \bigcap_{x\in Y} \lambda(x). 
$ 
For the CSA interpretation we refer to \secr{op_an}.
\ebox}
\end{lemma}

Combing this lemma with \lemr{tobs} for $G=G_1\synch G_2$, assuming that $\Sigma^{\veps}= \Sigma^{\veps}_1\, \dot\cup \,  \Sigma^{\veps}_2 \, \dot\cup \,   \Sigma^{\veps}_{12}$, where  $\Sigma^{\veps}_{12}$  includes the shared unobservable events in~$G_1$ and $G_2$  that become local after the synchronization $G_1\synch G_2$, we find that
\beq
\mc O_{\Sigma^{\veps}}(G_1 \synch G_2) = \mc O_{\Sigma^{\veps}_{12}}\big( \mc O_{\Sigma^{\veps}_1 \saa\dot\cup  \Sigma^{\veps}_2}(G_1\synch G_2)\big)=
\mc O_{\Sigma^{\veps}_{12}}\big(\mc O_{\Sigma^\veps_1}(G_1)\synch \mc O_{\Sigma^\veps_2}(G_2)\big)\eeq{eq:obsab}
Based on this result it is obvious that, in the case of unobservable shared events, the equality 
$\mc O_{\Sigma^\veps}(G_1 \synch G_2) = \mc O_{\Sigma^\veps_1}(G_1)\synch \mc O_{\Sigma^\veps_2}(G_2)$ does not always apply. This fact was also recently highlighted by an example in  \cite{Masopust_2018}. 

\subsection{Combined incremental observer generation and abstraction}
The challenge is now to combine the incremental observer generation in \rf{eq:obsab} with the incremental abstraction in Algorithm 1, here based on visible bisimulation since conflict equivalence is not applicable. If some unobservable events are shared and therefore cannot immediately be replaced by $\veps$, while some observable events are local and can be replaced by $\tau$ and then abstracted, the question is if it is possible to perform abstraction before  observer generation. The next example illustrates that this is not always possible. Before this example, two important remarks are given.
\b{enumerate}[(i)]
\item Initial local observers are always assumed to be generated before any hiding and abstraction. This means that every transition system $G$ in this section is by default an observer, although not explicitly expressed to simplify the notation. Thus, $G$ is assumed to be deterministic (except in the final \ther{theorem_imp}), and any non-safe states are labeled by $N$.

\item Hiding and abstraction are always performed on deterministic systems. Hence, alternative choices, including $\tau$ events after hiding, are interpreted as deterministic choices in observer generation. Restrictions will also be included such that repeated observer generation and abstraction (presented later in this section) still means that $G$ can be regarded as a deterministic transition system, although it may include alternative choices involving $\tau$ events.
\e{enumerate}

\fig{t}{   
\begin{tikzpicture}[->,xscale=0.9, yscale=0.9, automaton]
            \node at (-2.7em,1.3em) {$G$};
            \node[initial, state, label=$\{N\}$] (0) at (0,0) {$0$};
            \node[state, label=$\{N\}$] (1) at (5em,0)    {$1$};
            \node[state, label=$\{N\}$] (2) at (10em,0em)    {$2$};
            \node[state] (3) at (15em,0em)    {$3$};
            \path[->]
              (0) edge node [above] {$a$} (1)
              (1) edge node [above] {$b$} (2)
              (2) edge node [above] {$u$} (3);
\end{tikzpicture}
\yspa{1.3em}

\begin{tikzpicture}[->,xscale=0.9, yscale=0.9, automaton]
            \node at (-4.7em,1.3em) {$\mc O_{\{u\}}(G)$};
            \node[initial, state, rounded rectangle, label=$\{N\}$] (0) at (0,0) {$\{0\}$};
            \node[state, rounded rectangle, label=$\{N\}$] (1) at (5.5em,0)    {$\{1\}$};
            \node[state, rounded rectangle] (2) at (11.5em,0) {$\{2,3\}$};
            \path[->]
              (0) edge node [above] {$a$} (1)
              (1) edge node [above] {$b$} (2);
\end{tikzpicture}
\hspace{2.5em}
\begin{tikzpicture}[->,xscale=0.9, yscale=0.9, automaton]
            \node at (-6.6em,1.3em) {$\mc O_{\{u\}}(G)^{\mc A^{\{a,b\}}}$};
            \node[initial, state, rounded rectangle, label=$\{N\}$] (0) at (0,0) {$\{0,1\}$};
            \node[state, rounded rectangle] (1) at (6.5em,0)    {$\{2,3\}$};
            \path[->]
              (0) edge node [above] {$\tau$} (1);
\end{tikzpicture}
\yspa{1.4em}

\begin{tikzpicture}[->,xscale=0.9, yscale=0.9, automaton]
            \node at (-5.2em,1.3em) {$G^{\mc A^{\{a,b\}}}$};
            \node[initial, state, rounded rectangle, label=$\{N\}$] (0) at (0,0) {$\{0,1,2\}$};
            \node[state, rounded rectangle] (1) at (6.5em,0)    {$\{3\}$};
            \path[->]
              (0) edge node [above] {$u$} (1);
\end{tikzpicture}
\hspace{2.5em}
\begin{tikzpicture}[->,xscale=0.9, yscale=0.9, automaton]
            \node at (-9em,1.6em) {$\mc O_{\{u\}}(G^{\mc A^{\{a,b\}}})$};
            \node[initial, state, rounded rectangle] (0) at (0,0) {$\{\{0,1,2\},\{3\}\}$};
\end{tikzpicture}
}
{Different order between observer generation and abstraction for transition system $G$.}{f:abs_obs}            
            
\b{example}\label{ex:abs_obs}
Consider the deterministic transition system $G$ in \figr{f:abs_obs}, where the events $a$ and $b$ are observable, while $u$ is unobservable, and the first three states have label~$N$. In the observer $\mc O_{\{u\}}(G)^{\mc A^{\{a,b\}}}$, the unobservable event is first replaced by $\veps$ and the CSO observer is generated followed by visible bisimulation abstraction, while the observable events are first hidden and abstracted in $\mc O_{\{u\}}(G^{\mc A^{\{a,b\}}})$, followed by the CSO observer generation. 

State labels are preserved by visible bisimulation abstraction. This implies that abstracted block states with label $N$ only include states that before the abstraction were also labeled by $N$. Thus, observer generation before abstraction results in the correct solution, and the model $\mc O_{\{u\}}(G)^{\mc A^{\{a,b\}}}$ in \figr{f:abs_obs} shows that the states $0$ and $1$ are non-safe. 

Making the abstraction before the observer generation generates in this example a different and therefore incorrect result. The reason is that the states are then merged in wrong blocks in $G^{\mc A^{\{a,b\}}}$, where state $2$ is incorrectly joined with the states $1$ and~$0$. In the observer generation, where the non-safe block state $\{0,1,2\}$ is merged with the safe state $3$, the result is that the two states $0$ and $1$ incorrectly become safe. 
 \ebox 
 \e{example}

\paragraph{Avoiding abstractions that influence observer generation} 
Shared unobservable events mean that we need to repeat the observer generation when subsystems have been synchronized, since additional local unobservable events result in more $\veps$ transitions. At the same time, observable events must be abstracted when they become local, to avoid state space explosion. Thus, it is necessary to switch between abstraction and observer generation when subsystems are synchronized. According to \exr{ex:abs_obs} this is not possible without introducing some abstraction restrictions.

To avoid that any abstraction of a transition system influences later observer generations, additional state labels are added, assuming that some unobservable events have not yet been replaced by $\veps$. Unique state labels are then added to $\Sigma^{uo}$ source and target states.

\begin{definition}[\bm $\Sigma^{uo}$ source and target states]\label{d:uoST}
For a deterministic transition system~$G$, with a state set~$X$ and a set $\Sigma^{uo}$ of unobservable events, the states in the set
\[
X_{st}^{\Sigma^{uo}} = \{ x,x'\in X \st (\exists u\in {\Sigma_{uo}})\,x\trans{u}x'\}
\]
are called $\Sigma^{uo}$ {\em source and target states} (STSs).
\ebox
\end{definition}
Adding unique state labels for all $\Sigma^{uo}$ STSs means that visible bisimulation abstraction can be performed before observer generation. This is now illustrated for the transition system in \figr{f:abs_obs}. 

\b{example}\label{ex:uoST}
The same transition system $G$ as in Example \ref{ex:abs_obs} is considered, where unique state labels $\lambda^u_s$ and $\lambda^u_t$ are added to the source and target states of the transition $2\trans{u}3$. These $\Sigma^{uo}$ STS labels prevent the visible bisimulation abstraction  $G^{\mc A^{\{a,b\}}}$ to merge state $2$ with the states $0$ and $1$, since they have now different state labels. The succeeding CSO observer generation $\mc O_{\{u\}}(G)^{\mc A^{\{a,b\}}}$ merges the source and target states of the transition $2\trans{u}3$, and the obsolete $\Sigma^{uo}$ STS labels are removed. The resulting observer $\mc O_{\{u\}}(G)^{\mc A^{\{a,b\}}}$  in \figr{f:uoST} now coincides with the correct observer $\mc O_{\{u\}}(G)^{\mc A^{\{a,b\}}}$ in \figr{f:abs_obs}. 
\ebox 
\e{example}

\fig{t}{   
\begin{tikzpicture}[->,xscale=0.9, yscale=0.9, automaton]
            \node at (-2.7em,1.3em) {$G$};
            \node[initial, state, label=$\{N\}$] (0) at (0,0) {$0$};
            \node[state, label=$\{N\}$] (1) at (5em,0)   {$1$};
            \node[state, label={$\{N,\lambda^u_s\}$}] (2) at (10em,0em)    {$2$};
            \node[state, label=$\{\lambda^u_t\}$] (3) at (15em,0em)    {$3$};
            \path[->]
              (0) edge node [above] {$a$} (1)
              (1) edge node [above] {$b$} (2)
              (2) edge node [above] {$u$} (3);
\end{tikzpicture}
\yspa{1.6em}

\begin{tikzpicture}[->,xscale=0.9, yscale=0.9, automaton]
            \node at (-4.7em,1.3em) {$G^{\mc A^{\{a,b\}}}$};
            \node[initial, state, rounded rectangle, label=$\{N\}$] (0) at (0,0) {$\{0,1\}$};
            \node[state, label={$\{N,\lambda^u_s\}$}] (1) at (5.5em,0)   {$2$};
            \node[state, label=$\{\lambda^u_t\}$] (2) at (10.5em,0em)    {$3$};
            \path[->]
              (0) edge node [above] {$\tau$} (1)
              (1) edge node [above] {$u$} (2);
\end{tikzpicture}
\hspace{2.5em}
\begin{tikzpicture}[->,xscale=0.9, yscale=0.9, automaton]
            \node at (-6.7em,1.6em) {$\mc O_{\{u\}}(G^{\mc A^{\{a,b\}}})$};
            \node[initial, state, rounded rectangle, label=$\{N\}$] (0) at (0,0) {$\{0,1\}$};
            \node[state, rounded rectangle] (1) at (6.5em,0)    {$\{2,3\}$};
            \path[->]
              (0) edge node [above] {$\tau$} (1);
\end{tikzpicture}
}
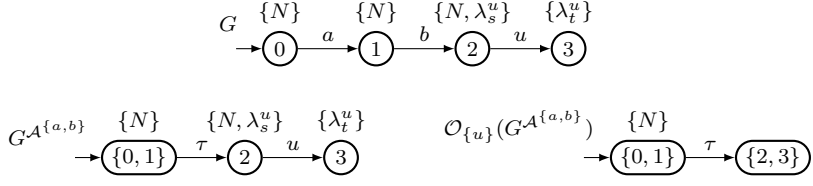
{Unique $\Sigma^{uo}$ STS labels $\lambda^u_s$ and $\lambda^u_t$ added to transition system $G$, resulting in a correct observer $\mc O_{\{u\}}(G^{\mc A^{\{a,b\}}})$ where visible bisimulation abstraction is performed before observer generation.}{f:uoST}            

\paragraph{Future nondeterministic choices}
Generally, an observer transforms a nondeterministic system to a language equivalent deterministic system. Based on Algorithm 1, the first operation generates an observer for every subsystem, which implies that any multiple initial states are replaced by one initial state in every initial observer, before any abstraction is made. Furthermore, nondeterministic choices are removed in the initial observer generation. However, additional nondeterministic choices may appear after additional $\veps$ transitions have been introduced. The following example illustrates this phenomenon and resulting complications.

\b{example}\label{x:nondet}
Consider the deterministic transition system $G$ in \figr{f:nondet}, where the events $a$, $b$ and $c$ are observable, while $u$ is unobservable. Additional $\Sigma^{uo}$ STS labels $\lambda^u_s$ and $\lambda^u_t$ are therefore also introduced in $G$. Making an abstraction before the observer generation implies that the event $a$ can not be replaced by $\tau$, since a future nondeterministic choice occurs when $u$ is replaced by $\veps$. Thus, consider the abstraction $G^{\mc A^{\{b,c\}}}$ in \figr{f:nondet} where $b$ and $c$ have been replaced by $\tau$, and the block state $\{3,4\}$ has been generated. Generating the CSO observer $\mc O_{\{u\}}(G^{\mc A^{\{b,c\}}})$, followed by one more abstraction $\mc O_{\{u\}}(G^{\mc A^{\{b,c\}}})^{\mc A^{\{a\}}}$, verifies that this observer coincides with the observer $\mc O_{\{u\}}(G)^{\mc A^{\{a,b,c\}}}$ that is always correct, since no abstraction is made before the observer generation.

This example indicates that abstraction can also be included before future nondeterministic choices have been removed by observer generation. Unfortunately, additional complications sometimes show up, in this example when event $c$ is replaced by event $b$ in $G$. Starting with visible bisimulation abstraction means that $G^{\mc A^{\{b\}}}$ and $\mc O_{\{u\}}(G^{\mc A^{\{b\}}})^{\mc A^{\{a\}}}$ coincide with $G^{\mc A^{\{b,c\}}}$ and $\mc O_{\{u\}}(G^{\mc A^{\{b,c\}}})^{\mc A^{\{a\}}}$, respectively, in \figr{f:nondet}. Thus, the observer where abstraction is involved before observer generation incorrectly includes a non-safe state, see \figr{f:nondett}, while the correct observer $\mc O_{\{u\}}(G)^{\mc A^{\{a,b\}}}$ in \figr{f:nondett} has no non-safe state.
 \ebox 
 \e{example}

\fig{t}{
         \begin{tikzpicture}[xscale=0.9, yscale=0.9, automaton]
            \node at (-2.5em,1.3em) {$G$};
            \node[state, label=$\{\lambda^u_s\}$, initial] (0) at (0em,0) {$0$};
            \node[state, label=$\{\lambda^u_t\}$] (1) at (5em,0) {$1$};
            \node[state] (2) at (10em,0) {$2$};
            \node[state, label=$\{N\}$] (3) at (15em,0) {$3$};
            \node[state, label=$\{N\}$] (4) at (20em,0) {$4$};
            \path[->]
              (0) edge  [out=-120,in=-60,loop,distance=3em] node [below] {$a$} (0)
              (0) edge  [out=30,in=150] node [above] {$u$} (1)
              (1) edge  [out=-150,in=-30] node [below] {$b$} (0)
              (1) edge node {$a$} (2)
              (2) edge node {$c$} (3)
              (3) edge node {$c$} (4);
           \end{tikzpicture}        
\yspa{0.4em}

         \begin{tikzpicture}[xscale=0.9, yscale=0.9, automaton]
            \node at (-4.2em,1.3em) {$G^{\mc A^{\{b,c\}}}$};
            \node[state, rounded rectangle, label=$\{\lambda^u_s\}$, initial] (0) at (0em,0) {$\{0\}$};
            \node[state, rounded rectangle, label=$\{\lambda^u_t\}$] (1) at (5em,0) {$\{1\}$};
            \node[state, rounded rectangle] (2) at (10em,0) {$\{2\}$};
            \node[state, rounded rectangle, label=$\{N\}$] (3) at (15em,0) {$\{3,4\}$};
            \path[->]
              (0) edge  [out=-120,in=-60,loop,distance=3em] node [below] {$a$} (0)
              (0) edge  [out=30,in=150] node [above] {$u$} (1)
              (1) edge  [out=-150,in=-30] node [below] {$\tau$} (0)
              (1) edge node {$a$} (2)
              (2) edge node {$\tau$} (3);
           \end{tikzpicture}        
\yspa{0.6em}

         \begin{tikzpicture}[xscale=0.9, yscale=0.9, automaton]
            \node at (-13.5em,1.3em) {$\mc O_{\{u\}}(G)^{\mc A^{\{a,b,c\}}}=\mc O_{\{u\}}(G^{\mc A^{\{b,c\}}})^{\mc A^{\{a\}}}$};
            \node[state, rounded rectangle, initial] (0) at (0em,0) {$\{0,1,2\}$};
            \node[state, rounded rectangle, label=$\{N\}$] (1) at (6.61em,0) {$\{3,4\}$};
            \path[->]
              (0) edge node {$\tau$} (1);
           \end{tikzpicture}        
}
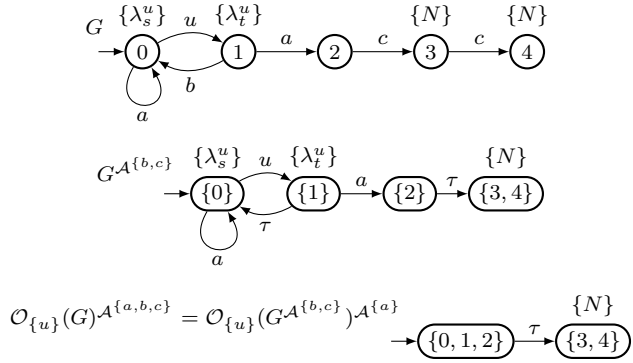
{Different order between observer generation and abstraction for  transition system $G$, including a nondeterministic choice when $u$ is replaced by $\veps$.}{f:nondet}

\fig{h}{
         \begin{tikzpicture}[xscale=0.9, yscale=0.9, automaton]
            \node at (-8.35em,1.3em) {$\mc O_{\{u\}}(G)^{\mc A^{\{a,b\}}}$};
            \node[state, rounded rectangle, initial] (0) at (0em,0) {$\{0,1,2,3,4\}$};
           \end{tikzpicture}        
\hspace{3em}
         \begin{tikzpicture}[xscale=0.9, yscale=0.9, automaton]
            \node at (-8.05em,1.3em) {$\mc O_{\{u\}}(G^{\mc A^{\{b\}}})^{\mc A^{\{a\}}}$};
            \node[state, rounded rectangle, initial] (0) at (0em,0) {$\{0,1,2\}$};
            \node[state, rounded rectangle, label=$\{N\}$] (1) at (6.61em,0) {$\{3,4\}$};
            \path[->]
              (0) edge node {$\tau$} (1);
           \end{tikzpicture}        
}
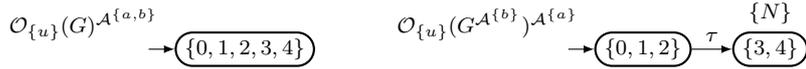
{Different order between observer generation and abstraction for transition system $G$ in \figr{f:nondet}, when the event $c$ is replaced by event $b$.}{f:nondett}

This example illustrates that even if an event label is preserved, which in the future will result in a nondeterministic choice, in this example event $a$, incorrect results may occur. Because correct results are often achieved, special rules can be established, but it is hard to define exactly when such rules give correct results. 

Our conclusion is therefore that abstractions of subsystems should not be performed before future nondeterministic choices have been removed by synchronization of subsystems and observer generation. Thus, it is necessary to exactly define a future nondeterministic choice.

\begin{definition}[\bm Future nondeterministic choice]\label{d:FNC}
For a deterministic transition system $G$ with a state set $X$, a set $\Sigma^o$ of observable events, and a set $\Sigma^{uo}$ of unobservable events, consider the set of transition relation sets
\bes
T_{nc}\eq \{(x,a,Y')\in X\times \Sigma^o\cup\Sigma^{uo} \times 2^X \st \yspa{0.05ex} \\ && (\exists x',x''\in Y')(\exists s_u,s_v\in (\Sigma^{uo}\setm\{a\})^*) \,  
 x\sa\trans{\xspa{0.5ex}s_ua\yspa{0.65ex}}x'\AND x\sa\trans{\xspa{0.5ex}s_va\yspa{0.65ex}}x''\AND x'\neq x''\,\}.
 \ees
The tuple $(x,a,x')$, where $x\in Y'$ and $(x,a,Y')\in T_{nc}$, is called a {\em future nondeterministic choice} (FNC) transition.
\ebox
\end{definition}

First, we observe that the complexity of finding FNC transitions is in worst case $O(|X||T|)$. However, the strings $s_u$ and $s_v$ are normally short, which means that worst case complexity is irrelevant. More importantly, the evaluation of FNC transitions is made on the original submodules $G_i$ in \rf{modular}, which are expected to be relatively small.

The exclusion of FNC transitions is only a minor restriction. First, note that unobservable identical events, not yet replaced by $\veps$, but generating FNC transitions, can without loss of generality be renamed to avoid such transitions that later nevertheless will be replaced by $\veps$ transitions. Observe that nondeterministic choice is the main reason for the well known worst case exponential state space complexity in observer generation. Thus, it is indeed recommended to accomplish this renaming, which of course also influences other submodels, including the same shared unobservable events.

For observable events, the expectation is that any nondeterministic choice is modeled explicitly, without adding  unobservable events before a nondeterministic choice (as event $u$ in \exr{x:nondet}). Explicit nondeterminism is then removed in the initial observer generation. Note that in this special case, where $s_u$ and $s_v$ are empty strings for $(x,a,Y')\in T_{nc}$, this relation can also be expressed by the function $Y'=\delta(x,a)$. 

To summarize, the assumption from now on is that all nondeterminism, except possible future $\veps$ transitions, is taken care of in the initial observer generation. 

\paragraph{Incremental  observer generation and abstraction}
Based on the definitions of $\Sigma^{uo}$ STSs and FNC transitions, we are now ready to prove \lemr{lemma:red}. This lemma states that a deterministic system including observer generation is equivalent to a system where abstraction is performed before the observer generation. The abstraction is assumed to be based on visible bisimulation.

We remind again that hiding and abstraction are always assumed to be performed on deterministic systems. Hence, alternative choices including $\tau$ events are interpreted as deterministic choices in observer generation. Also note that CSO and CSA observer generation only considers the non-safe $N$ labels, not the unique $\Sigma_{uo}$ STS labels, which are only included to avoid abstractions that influence the resulting state labels in the observer generation, see \figr{f:abs_obs}. When states in $\veps$ transitions are merged in the observer generation, related unique STS labels are also removed, since they are then obsolete. 

Due to the length of the following proof, an example is given after the proof where the relation between specific state sets and partitions is illustrated.

\begin{lemma}[Abstraction before observer generation]\label{lemma:red}
{\rm Consider a deterministic transition system $G= \langle X,\Sigma^o\cup\Sigma^{uo},T,I,AP,\lambda\rangle$ where $\Sigma^o$ is a set of observable events and $\Sigma^{uo}$ is a set of unobservable events. Let $\Sigma^h$ and $\Sigma^\veps$ be sets of local events that are replaced by $\tau$ and~$\veps$, respectively. Moreover, assume that unique state labels for all $\Sigma^{uo}$ STSs are included in $G$, while no FNC transition exists in $G$. Then
\[ 
\mc O_{\Sigma^\veps}(G)^{\Sigma^h} \sim \mc O_{\Sigma^\veps}(G^{\mc A^{\Sigma^h}}),
\]
where the abstraction is based on visible bisimulation.}

{Proof:} 
{\rm 
The state set $X$ is naturally divided into $X^o$, the set of states with observable transitions to and/or from states in $X^o$, and the set of $\Sigma^{uo}$ STSs $X^{\Sigma_{uo}}$, \ie $X= X^o{\mathbin{\dot{\cup}}} X^{\Sigma_{uo}}$. Note  that  border states in between observable and unobservable transitions belong to  $X^{\Sigma_{uo}}$. The main point of this proof is to show the separation between the abstraction, mainly acting on $X^o$, and the observer generation that only affects~$X^{\Sigma_{uo}}$. 

Before the observer generation, the unobservable events in $\Sigma^\veps$ are replaced by~$\veps$, and the transition relations in $T$ can then be divided into three parts:  $T^{\Sigma^o}$ including  observable transitions, $T^{\Sigma^\veps}$ only including $\veps$ transitions, and $T^{\Sigma^{uo}\setminus \Sigma^\veps}$ including remaining transitions with unobservable events, \ie $T=T^{\Sigma^o}{\mathbin{\dot{\cup}}} T^{\Sigma^\veps}{\mathbin{\dot{\cup}}} $ $T^{\Sigma^{uo}\setminus\Sigma^\veps}$. 

The observer generation $\mc O_{\Sigma^\veps}(\cdot)$ only affects the states in $X^{\Sigma_{uo}}$. The reason is that the transition function $G$ is deterministic before the introduction of $\veps$ transitions, and it does not include any FNC transitions. It means that for any choice of $\Sigma^\veps$, the transition relations in $T^{\Sigma^o}{\mathbin{\dot{\cup}}} T^{\Sigma^{uo}\setminus \Sigma^\veps}$ are all deterministic. More specifically, introducing the set
\bes
X^{\Sigma_{uo}}_{\mc O_{\Sigma^\veps}}\eq \{Y\in 2^X \st (\exists x\in X) (\exists a \in \Sigma^o{\mathbin{\dot{\cup}}}\Sigma^{uo}\setm\Sigma^\veps)(\exists x'\in X^{\Sigma_{uo}}) \\
        &&\;\;x\trans{a}x' \AND Y=R_\veps(x') \}
\ees
the observer state set can then be expressed as $X_{\mc O_{\Sigma^\veps}}=X^o{\mathbin{\dot{\cup}}} X^{\Sigma_{uo}}_{\mc O_{\Sigma^\veps}}$. Since the states in $X^{\Sigma_{uo}}_{\mc O_{\Sigma^\veps}}$ are block states, the states in $X^o$ are now also considered as singleton block states. The fact that the observer generation only affects the states in $X^{\Sigma_{uo}}$ is also valid for the observer state label updates for CSO and CSA. It is however critical that both the source and target states of $\veps$ transitions are included in $X^{\Sigma_{uo}}$, see \figr{f:abs_obs}. 

The structure of the observer state set $X_{\mc O_{\Sigma^\veps}}=X^o{\mathbin{\dot{\cup}}} X^{\Sigma_{uo}}_{\mc O_{\Sigma^\veps}}$ also implies that hiding of events for  observable transitions in $T^{\Sigma^o}$ can be performed either before or after the observer generation. The only difference between $T^{\Sigma^o}$ and the observer based version $T_{\mc O_{\Sigma^\veps}}^{\Sigma^o}$ is that any target states $x'\in X^{\Sigma_{uo}}$ for transitions in $T^{\Sigma^o}$ are replaced by block target states $Y\in X^{\Sigma_{uo}}_{\mc O_{\Sigma^\veps}}$ for transitions in $T_{\mc O_{\Sigma^\veps}}^{\Sigma^o}$. This does not influence the hiding mechanism that only changes the observable events in $\Sigma^h$ to $\tau$. Thus, $\mc O_{\Sigma^\veps}(G)^{\Sigma^h}=\mc O_{\Sigma^\veps}(G^{\Sigma^h})$. 

Now, consider the abstraction part, where the visible bisimulation partition $\Pi$ for $G^{\Sigma^h}$ is divided in the same way as the state set $X=X^o{\mathbin{\dot{\cup}}} X^{\Sigma_{uo}}$ such that $\Pi\define \Pi^o {\mathbin{\dot{\cup}}} \Pi^{\Sigma^{uo}}$. Since all states in $X^{\Sigma^{uo}}$ have unique state labels, corresponding block states in $\Pi^{\Sigma^{uo}}$ are singletons. This means that the abstraction only acts on the partition $\Pi^o$, resulting in possible non-singleton block states. Also note that the block states in $\Pi$ are the states of $G^{\mc A^{\Sigma^h}}$. 

The states $X_{\mc O_{\Sigma^\veps}}=X^o{\mathbin{\dot{\cup}}} X^{\Sigma_{uo}}_{\mc O_{\Sigma^\veps}}$ of the observer $\mc O_{\Sigma^\veps}(G^{\Sigma^h})$ only change the states in the unobservable part from $X^{\Sigma_{uo}}$ to $X^{\Sigma_{uo}}_{\mc O_{\Sigma^\veps}}$. Assume first that the unique STS labels still remain when $\veps$ transitions are merged in the observer generation. Every block state in $X^{\Sigma_{uo}}_{\mc O_{\Sigma^\veps}}$ then has a unique state label. This implies that the visible bisimulation partition after the observer generation can be expressed as $\Pi_{\mc O_{\Sigma^\veps}} = \Pi^o{\mathbin{\dot{\cup}}} X^{\Sigma_{uo}}_{\mc O_{\Sigma^\veps}}$, where once again the abstraction only acts on the partition $\Pi^o$. Thus, the abstraction only influences $X^o$ and generates the same partition $\Pi^o$ both with and without observer generation. The abstraction is therefore completely separated from observer generation.

Finally, the removal of the unique but obsolete $\Sigma^\veps$ STS labels after the observer generation generally results in a coarser partition $\Pi_{\mc O} \succeq \Pi_{\mc O_{\Sigma^\veps}}$. In the search for the coarsest partition $\Pi_{\mc O}$ the question is then if the same visible bisimulation partition is achieved starting from the state space $X_{\mc O_{\Sigma^\veps}}=X^o{\mathbin{\dot{\cup}}} X^{\Sigma_{uo}}_{\mc O_{\Sigma^\veps}}$ of the observer $\mc O_{\Sigma^\veps}(G)^{\Sigma^h}=\mc O_{\Sigma^\veps}(G^{\Sigma^h})$ or the state space $ \Pi_{\mc O_{\Sigma^\veps}} = \Pi^o{\mathbin{\dot{\cup}}} X^{\Sigma_{uo}}_{\mc O_{\Sigma^\veps}}$ of the observer $\mc O_{\Sigma^\veps}(G^{\mc A^{\Sigma^h}})$? Since all individual states of a block in a visible bisimulation partition have the same set of event-target-blocks according to \defr{dbs}, a coarser partition just means that more states have the same set of event-target-blocks. Thus, it is no restriction to start with a state space where the states in each block state have the same set of event-target-blocks as in $\Pi_{\mc O_{\Sigma^\veps}}$. A coarser partition $\Pi_{\mc O}$ only implies that some block states in $\Pi_{\mc O_{\Sigma^\veps}}$ will be merged, more exactly those that generate the same set of event-target-blocks when the obsolete $\Sigma^\veps$ STS labels are  removed, see further details in \cite{bl:des:2019} and \exr{ex:uoSTT}. 

To summarize, the visible bisimulation partition for $\mc O_{\Sigma^\veps}(G^{\mc A^{\Sigma^h}})$ is the same as for $\mc O_{\Sigma^\veps}(G)^{\Sigma^h}$ also for the coarser partition $\Pi_{\mc O}$, where the unique state labels for the $\Sigma^\veps$ STSs are removed, and therefore $\mc O_{\Sigma^\veps}(G)^{\Sigma^h} \sim \mc O_{\Sigma^\veps}(G^{\mc A^{\Sigma^h}})$.
\ebox}
\end{lemma}

\fig{b}{   
\begin{tikzpicture}[->,xscale=0.9, yscale=0.9, automaton]
            \node at (-2.7em,1.3em) {$G$};
            \node[initial, state, label=$\{N\}$] (0) at (0,0) {$0$};
            \node[state, label=$\{N\}$] (1) at (4.5em,0)   {$1$};
            \node[state, label=$\{\lambda^u_s\}$] (2) at (9em,0)   {$2$};
            \node[state, label={$\{N,\lambda^u_t\}$}] (3) at (13.5em,0)   {$3$};
            \node[state] (4) at (18em,0)   {$4$};
            \node[state, label=$\{\lambda^v_s\}$] (5) at (22.5em,0em)    {$5$};
            \node[state, label={$\{\lambda^v_t, \lambda^w_s\}$}] (6) at (27em,0em)    {$6$};
            \node[state, label=$\{\lambda^w_t\}$] (7) at (31.5em,0em)    {$7$};
            \node[state] (8) at (36em,0)   {$8$};
            \node[state] (9) at (40.5em,0)   {$9$};
            \path[->]
              (0) edge node [above] {$a$} (1)
              (1) edge node [above] {$b$} (2)
              (2) edge node [above] {$u$} (3)
              (3) edge node [above] {$a$} (4)
              (4) edge node [above] {$b$} (5)
              (5) edge node [above] {$v$} (6)
              (6) edge node [above] {$w$} (7)
              (7) edge node [above] {$a$} (8)
              (8) edge node [above] {$b$} (9);
\end{tikzpicture}
\yspa{1.8em}

\begin{tikzpicture}[->,xscale=0.9, yscale=0.9, automaton]
            \node at (-5.3em,1.3em) {$\mc O_{\Sigma^\veps}(G)^{\Sigma^h}$};
            \node[initial, state, rounded rectangle, label=$\{N\}$] (0) at (0,0) {$\{0\}$};
            \node[state, rounded rectangle, label=$\{N\}$] (1) at (4.8em,0)   {$\{1\}$};
            \node[state, rounded rectangle] (2) at (10.1em,0)   {$\{2,3\}$};
            \node[state, rounded rectangle] (3) at (15.4em,0)   {$\{4\}$};
            \node[state, rounded rectangle, label=$\{\lambda^w_s\}$] (4) at (20.7em,0em)    {$\{5,6\}$};
            \node[state, rounded rectangle, label=$\{\lambda^w_t\}$] (5) at (26em,0em)    {$\{7\}$};
            \node[state, rounded rectangle] (6) at (30.8em,0)   {$\{8\}$};
            \node[state, rounded rectangle] (7) at (35.6em,0)   {$\{9\}$};
            \path[->]
              (0) edge node [above] {$\tau$} (1)
              (1) edge node [above] {$\tau$} (2)
              (2) edge node [above] {$\tau$} (3)
              (3) edge node [above] {$\tau$} (4)
              (4) edge node [above] {$w$} (5)
              (5) edge node [above] {$\tau$} (6)
              (6) edge node [above] {$\tau$} (7);
\end{tikzpicture}
\yspa{0.82em}

\begin{tikzpicture}[->,xscale=0.9, yscale=0.9, automaton]
            \node at (-6.3em,1.3em) {$\mc O_{\Sigma^\veps}(G^{\mc A^{\Sigma^h}})$};
            \node[initial, state, rounded rectangle, label=$\{N\}$] (0) at (0,0) {$\{0,1\}$};
            \node[state, rounded rectangle] (1) at (6em,0)   {$\{2,3\}$};
            \node[state, rounded rectangle] (2) at (11.5em,0)   {$\{4\}$};
            \node[state, rounded rectangle, label=$\{\lambda^w_s\}$] (3) at (17em,0em)    {$\{5,6\}$};
            \node[state, rounded rectangle, label=$\{\lambda^w_t\}$] (4) at (22.5em,0em)    {$\{7\}$};
            \node[state, rounded rectangle] (5) at (28em,0)   {$\{8,9\}$};
            \path[->]
              (0) edge node [above] {$\tau$} (1)
              (1) edge node [above] {$\tau$} (2)
              (2) edge node [above] {$\tau$} (3)
              (3) edge node [above] {$w$} (4)
              (4) edge node [above] {$\tau$} (5);
\end{tikzpicture}
}
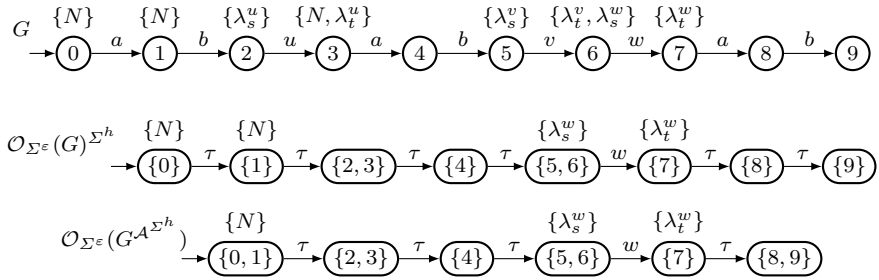
{Transition system $G$, corresponding observer $\mc O_{\Sigma^\veps}(G)^{\Sigma^h}$, and abstraction followed by observer $\mc O_{\Sigma^\veps}(G^{\mc A^{\Sigma^h}})$ for $\Sigma^h=\{a,b\}$ and $\Sigma^\veps=\{u,v\}$.}{f:lem_obs}

\b{example}\label{ex:uoSTT}
Consider the deterministic transition system $G$ in \figr{f:lem_obs} (assumed to be an observer), where $\Sigma^o=\Sigma^h=\{a,b\}$, $\Sigma^{uo}=\{u,v,w\}$, and $\Sigma^\veps=\{u,v\}$. 
Based on the notations in \lemr{lemma:red}, the involved state sets are $X^o=\{0,1,4,8,9\}$, $X^{\Sigma^{uo}}=\{2,3,5,6,7\}$, and $X^{\Sigma^{uo}}_{\mc O_{\Sigma^\veps}}=\{ \{2,3\},\{5,6\},\{7\} \}$. The involved visible bisimulation partitions are $\Pi^o=\{\{0,1\},\{4\},$ $\{8,9\}\},\Pi^{\Sigma^{uo}}=\{\{2\},\{3\},\{5\},\{6\},\{7\}\}$, and $\Pi^{\Sigma_{uo}}_{\mc O_{\Sigma^\veps}}=X^{\Sigma^{uo}}_{\mc O_{\Sigma^\veps}}$. The block states  in $\Pi_{\mc O_{\Sigma^\veps}} = \Pi^o{\mathbin{\dot{\cup}}} \Pi^{\Sigma_{uo}}_{\mc O_{\Sigma^\veps}}=\{\{0,1\},\{2,3\},\{4\},$ $\{5,6\},\{7\},\{8,9\}\}$ are the states of $\mc O_{\Sigma_\veps}(G^{\mc A^{\Sigma^h}})$, see \figr{f:lem_obs}. In the observer, the obsolete $\Sigma^\veps$ STS labels have been removed. Taking this into account, the visible bisimulation partitions for $\mc O_{\Sigma^\veps}(G)^{\Sigma^h}$ and $\mc O_{\Sigma^\veps}(G^{\mc A^{\Sigma^h}})$ are equal, both being $\Pi_{\mc O}=\{\{0,1\},\{2,3,4\},\{5,6\},$ $\{7\},\{8,9\}\}$. This confirms \lemr{lemma:red}, which says that these two observers are visible bisimulation equivalent.
\ebox 
\e{example}

Based on Lemmas \ref{tobs}-\ref{lemma:red} and \pror{prop:obs}, the following theorem shows how an abstracted observer for a modular system including shared unobservable events can be generated incrementally by combining observer generation and abstraction.

\begin{theorem}[Incremental observer generation and abstraction]\label{theorem_imp}
{\rm 
 Let $G_1$ and $G_2$ be two nondeterministic transition systems with hidden observable events in the set $\Sigma^h\define\Sigma^{h}_1 \saa\dot\cup \Sigma^{h}_2 \saa\dot\cup \Sigma^{h}_{12}$, where  $\Sigma^h_i$ includes local events in $G_i$, $i=1,2$, and $\Sigma^{h}_{12}$ includes shared events in $G_1$ and $G_2$. Furthermore, let $\Sigma^{uo}$ be the set of unobservable events for $G_1\synch G_2$, where $\Sigma^\veps\define\Sigma^{\veps}_1 \saa\dot\cup \Sigma^{\veps}_2 \saa\dot\cup \Sigma^{\veps}_{12}\subseteq \Sigma^{uo}$ and the events in $\Sigma^{\veps}_i$ are local for $G_i$, $i=1,2$, while the events in $\Sigma^{\veps}_{12}$ are shared events in $G_1$ and $G_2$. Also, assume that unique state labels for $\Sigma^{uo}\setm (\Sigma^{\veps}_1{\mathbin{\dot{\cup}}} \Sigma^{\veps}_2)$ STSs are included in $G_i$, while no FNC transition exists in $G_i$ for $i=1,2$. Then, a visible bisimulation abstraction of the observer $\mc O_{\Sigma^{\veps}}(G_1 \synch G_2)^{\Sigma^h}$ can be incrementally generated as}
\[
\mc O_{\Sigma^{\veps}}(G_1 \synch G_2)^{\Sigma^h} \sim \mc O_{\Sigma^{\veps}_{12}}( \mc O_{\Sigma^{\veps}_1}(G_1)^{\mc A^{\Sigma^h_1}} \synch \mc O_{\Sigma^{\veps}_2}(G_2)^{\mc A^{\Sigma^h_2}})^{\mc A^{\Sigma^h_{12}}}.
\]

{Proof:} 
{\rm  Based on \lemr{lemma:obs}, \rf{eq:obsab}, and $G=G_1\synch G_2$ in \lemr{lemma:red} we find that
\[
\mc O_{\Sigma^{\veps}}(G_1 \synch G_2)^{\Sigma^h_1 {\mathbin{\dot{\cup}}} \Sigma^h_2} \sim 
\mc O_{\Sigma^{\veps}_{12}}\big( (\mc O_{\Sigma^{\veps}_1}(G_1) \synch \mc O_{\Sigma^{\veps}_2}(G_2))^{\mc A^{\Sigma^h_1 {\mathbin{\dot{\cup}}} \Sigma^h_2}} \big).
\]
Together with \pror{prop:obs} and hiding of the events in $\Sigma^h_{12}$ on the left side, as well as an additional abstraction $\mc A^{\Sigma^h_{12}}$ on the right side, this gives
\[
\mc O_{\Sigma^{\veps}}(G_1 \synch G_2)^{\Sigma^h_1 {\mathbin{\dot{\cup}}} \Sigma^h_2 {\mathbin{\dot{\cup}}} \Sigma^h_{12}} \sim 
\mc O_{\Sigma^{\veps}_{12}}\big( \mc O_{\Sigma^{\veps}_1}(G_1)^{\mc A^{\Sigma^h_1}} \synch \mc O_{\Sigma^{\veps}_2}(G_2)^{\mc A^{\Sigma^h_2}}\big)^{\mc A^{\Sigma^h_{12}}},
\]
which proves the theorem.
\ebox}
\end{theorem}

The first local observer generations $\mc O_{\Sigma^{\veps}_i}(G_i)$, $i=1,2$, remove multiple initial states, non-deterministic choices and local unobservable events. The achieved observers guarantee that the following local hiding and abstraction is performed on deterministic systems. The critical point  investigated in \lemr{lemma:red}, including some minor restrictions, is that local abstractions can be accomplished before shared unobservable events are removed by the observer generation $\mc O_{\Sigma^{\veps}_{12}}$. This is followed by an additional abstraction ${\mc A^{\Sigma^h_{12}}}$ of the shared observable events that are only involved in $G_1\synch G_2$. 

This procedure is  repeated in the same way as in  Algorithm 1, which means that the state space explosion can be reduced significantly by the incremental abstraction, also for modular systems with shared unobservable events.

\paragraph{Algorithm} An algorithm for incremental observer generation and abstraction in the presence of shared unobservable events, presented in \ther{theorem_imp} for two subsystems, is generalized in the same way as in Algorithm 1. The only differences are that on line~2 unique state labels for source and target states of shared unobservable transitions are introduced after the observer generation, and line 8 is replaced with 
\[
G_\Omega:=\mc O ((G_{\Omega_1})^{\mc A} \synch (G_{\Omega_2})^{\mc A}). 
\]
In all abstractions, new local observable events are hidden,  and in all observer generations, new local unobservable events are replaced by $\veps$. After the observer generation on line 8, all obsolete state labels are also removed, as mentioned before \lemr{lemma:red}. 

Finally, note that no observer generation is necessary when there are no shared unobservable events in $G_{\Omega_1}$ and $G_{\Omega_2}$. A complement in the heuristics on selection of the sets $\Omega_1$ and $\Omega_2$ is therefore to also focus on subsystems that have shared unobservable events as early as possible. In this way, extra observer generations can be significantly reduced.

\section{Opacity verification of a multiple floor/elevator building} \label{sec:ex}
In order to demonstrate the practical use of the modular and incremental verification procedure, a CSO problem is formulated based on an $n$-story building with $m$ elevators on each floor. The model is inspired by an analytical and monolithic building example in \cite{dubreil10}. 

First, for better understanding of the problem, explicit results for the special case  $n=2$ and $m=2$ are presented in the following example.


\b{example}
Transition systems for a two-story building with floor models $F^1$ and $F^2$ and elevator models $E^1$ and $E^2$ are shown in \figr{exf2e2}. Each floor consists of two corridors and two elevators. Elevator entrances are located in states $2$ and $4$ in $F^i$ and $E^j$ for $i,j=1,2$. There are card readers in the corridors and elevators, which are represented by the events $c^i_j$ and $e^i_j$, respectively. The subscript $j$ indicates the corridor and elevator and the superscript $i$ corresponding floor. The events $u_j$ and $d_j$ indicate the upward and downward movement of the $j$-th elevator, respectively. The shared elevator event $e^i_j$ coordinates floor $i$ with elevator $j$, and the alternative choices with equal events, $c^i_1$ in state $2$ and $c^i_2$ in state $4$, plus arbitrary initial states ($I^{F^i}=X^{F^i}$ and $I^{E^j}=X^{E^j}$) generate the non-determinism of the system.

The observer of the $i$-th floor $\mc O(F^i)$ is also shown in \figr{exf2e2}, where opacity depends on the choice of secret states. Assume that there are only secret states in the second floor model $F^2$, which means that there can only be non-safe states in the observer $\mc O(F^2)$. The opacity of the total system $F_1\synch F_2\synch E_1\synch E_2\synch$ is therefore determined by the state labels of $\mc O(F^2)$. If the secret state is $X^{F^2}_S=\{1\}$ or $X^{F^2}_S=\{3\}$, the system is opaque, since all block states in  $\mc O(F^2)$ then include non-secret states, meaning that all states are safe. On the other hand, $X^{F^2}_S=\{1,3\}$ results in a non-opaque system, since the state $\{1,3\}$ in $\mc O(F^2)$ then becomes non-safe.
\ebox
\e{example}

\fig{t}{\begin{tikzpicture}[xscale=1, yscale=1, automaton]
            \node at (-6.8em,5.5em) {$F^i$};
            \node[state] (1) at (-5em,4em) { $1$};
            \node[state] (2) at (-1em,4em) { $2$};
            \node[state,rounded rectangle] (22) at (3em,4em) { $2'$};
            \node[state,rounded rectangle] (4) at (-5em,0) { $4$};
            \node[state,rounded rectangle] (44) at (-9em,0em) { $4'$};
            \node[state,rounded rectangle] (3) at (-1em,0) {{$3$}};
            \path[->]
              (1) edge [] node {$c^i_1$} (2)
              (2) edge [] node {}        (1)
              (2) edge [] node {$c^i_1$} (3)
              (2) edge [] node {$e^i_1$} (22)
              (22) edge [] node {} (2)
              (4) edge [] node {$c^i_2$} (1)
              
              (4) edge [] node {$e^i_2$} (44)
              (44) edge [] node {} (4)
              (4) edge [] node {$c^i_2$} (3)
              (3) edge [] node {} (4);
           \end{tikzpicture}
            \hspace{4.7em}
         \begin{tikzpicture}[xscale=1, yscale=1, automaton]
            \node at (-2.3em,1.4em) {\xspa{0.8em}$E^j$};
            \node[state] (0) at (0,0)    {$1$};
            \node[state] (00) at (0,-4em)    {$2$};
            \node[state] (1) at (5em,0)  {$3$};
            \node[state] (11) at (5em,-4em)  {$4$};
            \path[->]
              (0) edge [out=30,in=150] node [above] {$u_j$} (1)
              (0)  edge [] node [right] {$e^1_j$} (00)
              (00) edge [] node [above] {} (0)
              (1) edge [out=-150,in=-30] node [above] {$d_j$} (0)
              (1)  edge [] node [right] {$e^2_j$} (11)
              (11) edge [] node [above] {} (1);
           \end{tikzpicture}
           \begin{tikzpicture}[xscale=1, yscale=1, automaton]
\node[state,above,initial above,initial text={$\mathcal O(F^i)$},minimum size=1ex,rounded rectangle] (init) at (0em,-.25em)  {$\{1,2,3,4,2',4' \}$};
\node[state,minimum size=1ex,rounded rectangle]    (123)  at (9.5em,.5em)    {$\{1,2,3\}$};
\node at (4em,-1.4em) {\xspa{0.8em}$c^i_2$};
\node at (-5em,-1.4em) {\xspa{0.8em}$c^i_1$};
\node[state,minimum size=1ex,rounded rectangle]    (134)  at (-9.5em,.5em)   {$\{1,3,4\}$};
\node[state,minimum size=1ex,rounded rectangle]    (25)   at (4em,-5em)   {$\{2,2'\}$};
\node at (5.8em,-3.0em) {\xspa{0.8em}$e^i_1$};
\node[state,minimum size=1ex,rounded rectangle]    (46)   at (-4em,-5em)  {$\{4,4'\}$};
\node at (-6.5em,-3.0em) {\xspa{0.8em}$e^i_2$};
\node[state,minimum size=1ex,rounded rectangle]    (5)    at (9.5em,-5em)  {$\{2'\}$};
\node[state,minimum size=1ex,rounded rectangle]    (6)    at (-9.5em,-5em) {$\{4'\}$};
\node[state,minimum size=1ex,rounded rectangle]    (2)    at (9.5em,-10em)   {$\{2\}$};
\node[state,minimum size=1ex,rounded rectangle]    (4)    at (-9.5em,-10em)  {$\{4\}$};
\node[state,minimum size=1ex,rounded rectangle]    (13)   at (0em,-10em)  {$\{1,3\}$};
\path[->] 		
(init) edge [] node [above] {$c^i_1$} (123)
	   edge [] node [above] {$c^i_2$} (134)
       edge [] node [above] {\xspa{0.3em}$e^i_1$} (25)
       edge [] node [above] {$e^i_2$\xspa{0.5em}} (46)
(123) edge [out=-20,in=30, loop,distance=2.8em] node [right] {$c^i_1$} (123)
      edge [] node [right] {$e^i_1$} (5)
      edge [] node [below] {} (4)
(134) edge [out=200,in=150, loop,distance=2.8em] node [left] {$c^i_2$} (134)
      edge [] node [left] {$e^i_2$} (6)
      edge [] node [below] {} (2)
(5)  edge [] node [right] {$e^i_1$} (2)
(2)  edge [] node  {} (5)
(6)  edge [] node [left] {$e^i_2$} (4)
(4)  edge [] node  {} (6)
(46) edge [out=210,in=150,loop,distance=2.8em] node  {} (46)
     edge [] node [below]  {$c^i_2$\xspa{0.3em}} (13)
(25) edge [out=-30,in=30,loop,distance=2.8em] node  {} (25)
     edge [] node [below] {$c^i_1$} (13)
(4) edge [] node  {} (13)
(2) edge [] node  {$c^i_1$} (13)
(13) edge [] node  {} (2)
(13) edge [] node  {$c^i_2$} (4);
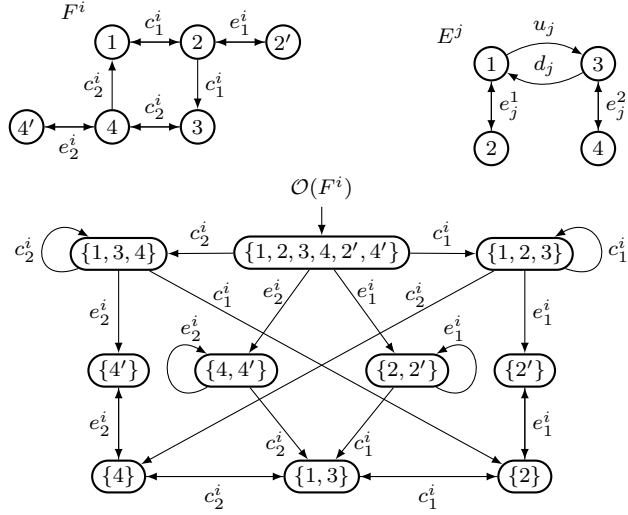
\end{tikzpicture}}{Floor plan model $F^i$ for the $i$-th floor, elevator model $E^j$ for the $j$-th corridor $c_j$, and observer $\mathcal O(F^i)$ for the $i$-th floor.}{exf2e2}

\paragraph{Building with $n$ floors and $m$ elevators} Consider an $n$-story building with $m$ elevators on each floor. The transition system models of floors and elevators are depicted in \figr{ex_FE}. There are $n$ floor models $F^i$, $i\in\mathbb{N}^+_n$ and $m$ elevator models  $E^j$, $j\in\mathbb{N}^+_m$. Each floor consists of corridors, rooms and elevators. Elevator entrances are located in states $2,4,\dots,2m-2,2m$ in $F^i$ and states $2,4,\dots,2n-2,2n$ in $E^j$. Corridors are connected through doors that open using card readers. The card readers are installed at the entrances of the elevators. Passing through the entrances of corridors and elevators are shown by events $c$ and $e$, respectively. Note that $c^i_j$ ($e^i_j$) indicates the $j$-th corridor (elevator) on the $i$-th floor. The events $u_j$ and $d_j$ represent the upward and downward movement of the $j$-th elevator, respectively. 

The shared elevator event $e^i_j$ coordinates floor $i$ with elevator $j$, and in state $2j$ on floor $i$ an alternative choice with equal event, $c^i_j$, as well as arbitrary initial states  generate the system non-determinism. The staff moving patterns can be tracked by observing the records of their ID cards that are read by the card readers. All floors have similar structure, but have different secret states which are places in the building that have a storage for secret documents.

\fig{t}{\begin{tikzpicture}[xscale=1, yscale=1, automaton]
            \node at (-6.7em,6.4em) {$F^i$};
            \node[state] (1) at (-5em,4.4em) { $1$};
            \node[state] (2) at (0em,4.4em) { $2$};
            \node[state] (22) at (0em,9em) { $2'$};
            \node[state] (3) at (5em,4.4em) { $3$};
            \node[state] (4) at (10em,4.4em) { $4$};
            \node[state] (44) at (10em,9em) { $4'$};
            \node[] (5) at (15em,4.4em) {\ldots};
            \node[state,rounded rectangle] (6) at (-5em,0) { ${2m}$};
            \node[state,rounded rectangle] (66) at (-5em,-4.6em) { ${2m'}$};
            \node[state,rounded rectangle] (7) at (1.5em,0) {{${2m-1}$}};
            \node[state,rounded rectangle] (8) at (8.75em,0) {${2m-2}$};
            \node[state,rounded rectangle] (88) at (8.75em,-4.6em) {${2m-2}'$};
            \node[state,rounded rectangle] (9) at (16em,0) {${2m-3}$};
            \node[]      (10) at (22em,0) {$\ldots$};
            \path[->]
              (1) edge [] node {$c^i_1$} (2)
              (2) edge [] node {}        (1)
              (2) edge [] node {$c^i_1$} (3)
              (2) edge [] node {$e^i_1$} (22)
              (22) edge [] node {} (2)
              (3) edge [] node {$c^i_2$} (4)
              (4) edge [] node {} (3)
              (4) edge [] node {$e^i_2$} (44)
              (44) edge [] node {} (4)
              (4) edge [] node {$c^i_2$} (5)
              (6) edge [] node {$c^i_m$} (1)
              (6) edge [] node {$e^i_m$} (66)
              (66) edge [] node {} (6)
              (6) edge [] node {$c^i_m$} (7)
              (7) edge [] node {} (6)
              (7) edge [] node {$c^i_{m-1}$} (8)
              (8) edge [] node {$c^i_{m-1}$} (9)
              (8) edge [] node {$e^i_{m-1}$} (88)
              (88) edge [] node {} (8)
              (9) edge [] node {} (8)
              (9) edge [] node {$c^i_{m-2}$} (10);
           \end{tikzpicture}
         \yspa{1.3em}
         \begin{tikzpicture}[xscale=1, yscale=1, automaton]
            \node at (-2em,2em) {\xspa{0.8em}$E^j$};
            \node[state] (0) at (0,0)    {$1$};
            \node[state] (00) at (0,-4.6em)    {$2$};
            \node[state] (1) at (5em,0)  {$3$};
            \node[state] (11) at (5em,-4.6em)  {$4$};
            \node[]      (2) at (10em,0) {$\dots$};
            \node[state,rounded rectangle] (3) at (15em,0) {$2n-1$};
            \node[state,rounded rectangle] (33) at (15em,-4.6em) {$2n$};
            \path[->]
              (0) edge [out=30,in=150] node [above] {$u_j$} (1)
              (0)  edge [] node [right] {$e^1_j$} (00)
              (00) edge [] node [above] {} (0)
              (1) edge [out=-150,in=-30] node [above] {$d_j$} (0)
              (1) edge [out=30,in=150] node [above] {$u_j$} (2)
              (1)  edge [] node [right] {$e^2_j$} (11)
              (11) edge [] node [above] {} (1)
              (2) edge [out=-150,in=-30] node [above] {$d_j$} (1)
              (2) edge [out=30,in=160] node [above] {$u_j$} (3)
              (3)  edge [] node [right] {$e^n_j$} (33)
              (33) edge [] node [above] {} (3)
              (3) edge [out=-150,in=-40] node [above] {$d_j$} (2);
           \end{tikzpicture}
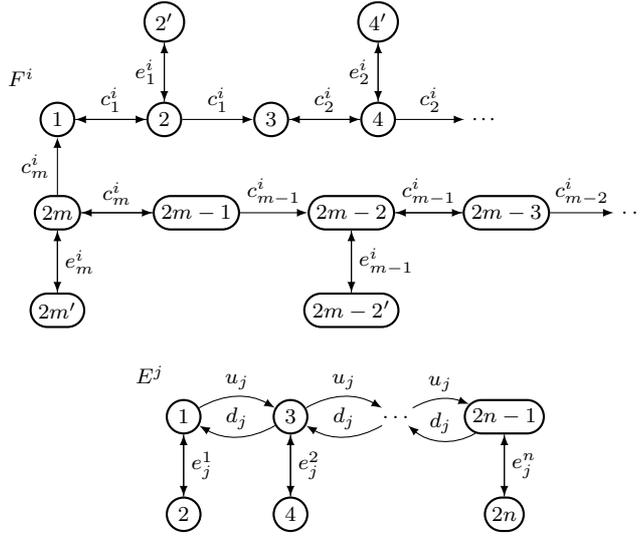
\vspace{0.25em}}{Floor plan model $F^i$ for the $i$-th floor and elevator model $E^j$ for the $j$-th corridor $c_j$.}{ex_FE}

\paragraph{Two scenarios}
One member of the staff wants to place a secret document in one of the secret locations in the building. There is an intruder that knows the structure of the system and has access to the records of card readers. The question is then if the intruder can have the knowledge that the secret document is in that specific location. In this case, an opaque system means that even a very careless staff with no specific strategy can place the secret document in any of the secret places, without being concerned of the secret being revealed. Most opacity examples have only one strategy for reaching to an opaque solution, while in this example, no specific strategy is needed as long as the system is opaque.

As an alternative scenario, consider the case when some card readers do not work due to power failure. Since corresponding doors  in that case cannot be opened, related transitions are then removed.

\paragraph{Results}
The CSO verification results for the building in \figr{ex_FE}, with different number of floors and elevators, are presented in \tabr{tab:comp:trans} for non-opaque systems. Results for the alternative scenario, where some doors can not be opened due to power failure, are shown in \tabr{tab:comp:opaq}. Restrictions are then introduced such that all systems become opaque.

In the first column of both tables, the pair $(n,m)$ shows the number of floors and elevators. The second column in \tabr{tab:comp:trans} includes the set of secret states on each floor. The set of secret states in \tabr{tab:comp:opaq} are the same as in \tabr{tab:comp:trans}, but omitted due to shortage of space. The column in both tables that is indicated with a $\star$ sign, shows the floors with local non-safe states, and in \tabr{tab:comp:opaq} also the corridors where card readers do not work.

The number of states  $|\what X|$ and transitions $|\what T|$, as well as the elapsed time for the verification $t_e$, are then presented for observers with abstraction $\mathcal O(G)^{\mathcal A}$ and without abstraction $\mathcal O(G)$. The verification is based on the transformation to a nonblocking problem, including conflict equivalence abstraction, as presented in \secr{s:extobs}.

\tab{t}{Results for non-opaque systems, where the number of states  and transitions  plus execution time are given for observers with abstraction $\mathcal O(G)^{\mathcal A}$ and without abstraction $\mathcal O(G)$. The $\star$ sign column shows the floors with local non-safe states.\yspa{1.4ex}\hfill}
{\begin{tabular}{ccc|rrr|rrr}\toprule
\multicolumn{3}{ c }{ \yspa{1.9ex} $F^i/E^j$-model} & \multicolumn{3}{ |c| }{$\mathcal O(G)^{\mathcal A}= \parallel \mathcal O(G_i)^{\mathcal A}$ \yspatb{0.8ex}{1ex}} & \multicolumn{3}{ c }{$\mathcal O(G)=\parallel \mathcal O(G_i)$} \\ \hline
$(n,m)$ & $X^i_s$ & $\star$ & \yspatb{1.2ex}{1.2ex} $|\what X|$ & $|\what T|$ &$t_e$\,(ms) & $|\what X|$ & $|\what T|$ &$t_e$\,(ms)\yspatb{0.6ex}{1ex} \\ \midrule
(1,1) & \yspatb{0.5ex}{1.7ex}$X^{F^1}_S=\{1\}$&$F^1$ &9 &13&2&7&11&8 \\ \cline{2-2}
(1,3) & \yspatb{0.8ex}{1.7ex}$X^{F^1}_S=\{1,2,3\}$ & $F^1$&26&50&4&19&39&9\\ \cline{2-2}	
(2,2) & \makecell{\yspatb{0.8ex}{1.5ex}$X^{F^1}_S=\{3\}$ \\ \yspa{1.5ex} $X^{F^2}_S=\{1,5\}$} &  $F^2$   & 49 & 127 &  5 & 775&2,913&15  \\ \cline{2-2}
(2,3) & \makecell{\yspatb{0.8ex}{1.5ex}$X^{F^1}_S=\{3\}$\\ \yspa{1.5ex} $X^{F^2}_S=\{1,5\}$} &  $F^2$ &25&52& 5  & 3,823 & 14,567& 44 \\\cline{2-2}
(3,3) & \makecell{\yspatb{0.8ex}{1.5ex}$X^{F^1}_S=\{1\}$ \\ \yspa{1.5ex} $X^{F^2}_S=\{1,3\}$\\ \yspa{1.5ex} $X^{F^3}_S=\{1,5\}$ } & \makecell{\yspa{1.5ex} $F^2$\\ \yspa{1.5ex} $F^3$}  & 32 & 89 & 6  & 347,445 &1,732,836 & 10,918  \\ \cline{2-2}
(3,4) & \makecell{\yspatb{0.8ex}{1.5ex}$X^{F^1}_S=\{1\}$\\ \yspa{1.5ex} $X^{F^2}_S=\{1,3\}$\\ \yspa{1.5ex} $X^{F^3}_S=\{5,7\}$} & \makecell{\yspa{1.5ex} $F^2$ \\ \yspa{1.5ex} $F^3$} & 42&118&7&$\approx 2.6 \cdot 10^6$ & o.m.  & -- \\\cline{2-2}
(4,3) & \makecell{\yspatb{0.8ex}{1.5ex}$X^{F^1}_S=\{1\}$ \\ \yspa{1.5ex} $X^{F^2}_S=\{1,3\}$ \\ \yspa{1.5ex} $X^{F^3}_S=\{1,5\}$\\ \yspa{1.5ex} $X^{F^4}_S=\{1,2,3\}$}&\makecell{\yspa{1.5ex} $F^2$\\ \yspa{1.5ex} $F^3$\\ \yspa{1.5ex} $F^4$}&32&105& 4& $\approx 16 \cdot 10^6$ & o.m. & --  \\ \cline{2-2}
(5,3) & \makecell{\yspatb{0.8ex}{1.5ex}$X^{F^2}_S=\{1,3\}$\\ \yspa{1.5ex} $X^{F^3}_S=\{1,5\}$ \\ \yspa{1.5ex} $X^{F^4}_S=\{1,2,3\}$\\ \yspa{1.5ex} $X^{F^5}_S=\{3,4,5\}$}&\makecell{\yspatb{0.8ex}{1.5ex} $F^2$\\ \yspa{1.5ex} $F^3$\\ \yspa{1.5ex} $F^4$\\ \yspa{1.5ex} $F^5$} &32&121&9& $\approx 22.6 \cdot 10^6$ &o.m. &-- \\ \bottomrule
\end{tabular}}{tab:comp:trans}
\tab{t}{Results for opaque systems, where number of states  and transitions  plus execution time are given for observers with abstraction $\mathcal O(G)^{\mathcal A}$ and without abstraction $\mathcal O(G)$. The $\star$ sign column shows the corridors where card readers do not work.\yspa{1.4ex}\hfill}
{\begin{tabular}{cc|rrr| rrr} \toprule
 \multicolumn{2}{ c }{\yspa{1.9ex} $F^i/E^j$-model} & \multicolumn{3}{| c| }{$\mathcal O(G)^{\mathcal A}=\parallel \mathcal O(G_i)^{\mathcal A}$ \yspatb{0.8ex}{1ex}} & \multicolumn{3}{ c }{$\mathcal O(G) =\parallel \mathcal O(G_i)$} \\ \hline
$(n,m)$  & $\star$ & \yspatb{1.2ex}{1.2ex} $|\what X|$ & $|\what T|$ &$t_e$\,(ms) & $|\what X|$ & $|\what T|$ &$t_e$\,(ms)\yspatb{0.6ex}{1ex} \\\midrule
(1,1) & \yspatb{0.5ex}{1.7ex}  $(c^1_{1})$ & 2 & 1 & 1 & 2 & 2 & 7\\  
(1,3) & \yspa{1.7ex}  $(c^1_{1})$ & 12 & 19 & 3 & 12 & 24 & 8\\  
(2,2) & \yspa{1.7ex}  $(e^2_{1})$  & 13 & 30 & 7  & 487 & 2,014 & 12\\  
(2,3) & \yspa{1.7ex}  $(c^2_{3})$& 44 & 112 &  10 & 1,699  & 6,450 & 24\\ 
(3,3) &  \yspa{1.7ex} ($c^2_{1}$, $c^3_{3})$  & 266 & 1,048 & 45 & 75,520 & 380,508 & 1,673\\  
(3,4) &   \yspa{1.7ex} $(c^2_{1}$, $c^3_{3})$ &413&1,667&106 & 591,867 & 3,042,099 & 18,058 \\ 
(4,3) &   \yspa{1.7ex} $(c^2_{1}$, $c^3_{3}$, $c^4_{1})$ &3,043&17,807&10,530& $\approx 2 \cdot 10^6$ &o.m. &--\\
(5,3) &\yspa{1.5ex} $(c^2_{1}$, $c^3_{3}$, $c^4_{1}$, $c^5_{2})$ & 19,556 & 142,087 & 24', 39'' & $ \approx 23.1 \cdot 10^6$ &o.m.&-- \\ \bottomrule
\end{tabular}}{tab:comp:opaq}
The results in Tables~\ref{tab:comp:trans} and \ref{tab:comp:opaq}  clearly demonstrate the strength of including the incremental abstraction. The number of states in \tabr{tab:comp:opaq} for three floors and four elevators is more than 1000 times larger when abstraction is not included, and no solution is obtained without abstraction for the larger systems. We also observe that for the non-opaque case in \tabr{tab:comp:trans} the computation does not continue when one of the non-safe states has been found, which makes it much faster than the verification of opaque systems, where the whole reachable state space (although abstracted) needs to be evaluated.

\section{Opacity and anonymity enforcement}
For a transition system $G$, with non-safe block states defined for CSO in \defr{d:cso} and CSA in \defr{d:csa}, it will now be shown how opacity/anonymity can be enforced by a supervisor $S$. Only safe block states will then be reachable in the controlled (closed loop) system. Two assumptions that simplify the computation of the supervisor are introduced: 
\begin{enumerate}[(i)]
\item All unobservable events are assumed to be uncontrollable. Since only observable events are then controllable and therefore can be disabled by the supervisor, it can be generated by the observer $\mc O(G)$ that is relevant for the actual security problem. The assumption is the same as in \cite{saboori:2012}, but a simpler CSO formulation is presented here.
\item All unobservable events are assumed to be local. This means that  the observer $\mc O(G)$ of a modular system \rf{modular} can be obtained by synchronizing local observers~\rf{obs_eq}, and efficient abstractions can be used in the supervisor synthesis.
\end{enumerate}
The input to the supervisor $S$ from the transition system $G$ only includes observable events, and $S$ restricts the behavior of the controlled system by disabling some controllable events \citep{rw:con:1989, kg:mod:1995}. Since all controllable events here are also assumed to be observable, the information to and from the supervisor only involves observable events. Thus, the nondeterministic transition system $G$ is perfectly represented by its observer $\mc O(G)$ in the synthesis and implementation of the supervisor $S$. This means that the closed loop system, where $S$ is also represented as an automaton, can be modeled as $\mc O(G)\synch S$. 

\subsection{Observer based supervisor generation}
The non-safe block states in $G$ that must be avoided by the supervisor are determined by the observer $\mc O(G)$, according to Props.~\ref{p:cso} and \ref{p:csa}. These states are now called {\em forbidden states} and included in the set $\what X_f$. Since no other state labels are considered, the observer is simplified from an arbitrary transition system to the automaton $\mathcal O (G) = \langle \what X,\Sigma,\what T,\what  I \rangle$. Furthermore, a restricted observer automaton is introduced 
\[
\mc O(G)_{\setminus \what X_f}=\langle\what X\setm \what X_f , \Sigma, \what T_{\setminus \what X_f} ,\what I\setm \what X_f \rangle
\]
where $\what T_{\setminus \what X_f} = \{(Y,a,Y') \smm\in \smm\what T \st  Y, Y' \notin \what X_f\}$. Hence, $\mc O(G)_{\setminus \what X_f}$  is the automaton where all forbidden states are excluded. 

If all events in $\mc O(G)$ are controllable, the automaton $S=\mc O(G)_{\setminus \what X_f}$ is the maximally permissive supervisor for the closed loop system $\mc O(G) \synch S$. When also some observable transitions are uncontrollable, these events can not be disabled by the supervisor. Not only the forbidden states must then be excluded in $S$, but also states from which there are uncontrollable transitions to forbidden states. The following proposition shows how the supervisor is then obtained by generating an extended set of forbidden states $\what X^e_f$. It is a minor modification and special case of the supervisor synthesis method presented in \cite{mf:csdes:2008}.

\begin{proposition}[Observer Based Supervisor] \label{pro:cce}  \xspa{0.1em}
{\rm
Consider a transition system $G$ and corresponding observer $\mc O(G)$ with state space $\what X$, uncontrollable event set $\Sigma_u$,  transition function $\what \delta$, and forbidden state set $\what X_f$. The {\em extended forbidden state set} 
\[
\what X^e_{f} = \{ Y \in\what X \st  (\exists s_u\in \Sigma^*_u) \, \what\delta(Y,s_{u})\in \what X_f \}
\]
results in the maximally permissive and controllable supervisor $S=\mc O(G)_{\setminus \what X^e_f}$ \lb

\vspace{-1.2em} 
\noindent for the closed loop system $\mc O(G)\synch S=S$. 
}

{Proof (sketch):} {\rm Since the supervisor $S=\mc O(G)_{\setminus \what X^e_f}$ is a subautomaton of $\mc O(G)$ and both are deterministic, it follows that the supervisor is a model of the closed loop system, \ie $\mc O(G)\synch S=S$. In the supervisor, only those states are removed from which it is possible to reach a forbidden state by only executing uncontrollable transitions. Thus, the supervisor $S$ is maximally permissive. Furthermore, all \text{border} transitions entering $\what X^e_{f}$ are controllable, and since these transitions are removed from~$S$, only controllable events are disabled. Hence, the supervisor is controllable. 
\ebox}
\end{proposition}

\fig{b}{
        \begin{tikzpicture}[xscale=0.85, yscale=0.85, automaton]
            \node at (-2.9em,1.5em) {$G$};
            \node[state, initial] (0) at (0,0) {$0$};
            \node[state, initial above] (1) at (6.5em,0) {$1$};
            \node[state] (2) at (13em,0) {$2$};
            \node[state] (3) at (0,-6.5em) {$3$};
            \node[state] (4) at (6.5em,-6.5em) {$4$};
            \node[state] (5) at (13em,-6.5em) {$5$};
           \path[->]
              (0) edge [out=120,in=60,loop,distance=2.5em] node [above] {$a$} (0)
              (0) edge node {$a$} (1)
              (1) edge node {$\veps$} (2)
              (0) edge node [left] {$c$} (3)
              (3) edge node {$\veps$} (4)
              (4) edge node {$!d$} (5)
              (0) edge [out=-35,in=-145] node [below] {\xspa{0.2em}$b$} (2);
         \end{tikzpicture} \xspa{3em}       
        \begin{tikzpicture}[xscale=0.85, yscale=0.85, automaton]
            \node at (-4.7em,1.5em) {$\mc O(G)$};
            \node[state, initial, inner sep=1.4pt, rounded rectangle] (0) at (0,0) {$\{0,\!1,\!2\}$};
            \node[state, inner sep=1.4pt, rounded rectangle, label=above: $\{N\}$] (1) at (8em,0) {$\{2\}$};
            \node[state, inner sep=1.4pt, rounded rectangle] (2) at (0,-6.5em) {$\{3,\!4\}$};
            \node[state, inner sep=1.4pt, rounded rectangle, label=above: $\{N\}$] (3) at (8em,-6.5em) {$\{5\}$};
           \path[->]
              (0) edge [out=120,in=60,loop,distance=2.5em] node [above] {$a$} (0)
              (0) edge node {$b$} (1)
              (0) edge node [left] {$c$} (2)
              (2) edge node {$!d$} (3);
         \end{tikzpicture}   \xspa{1em}     
}{Transition system $G$ in \exr{ex:synth} and its observer $\mathcal O(G)$.}{f:synth}

\b{example}\label{ex:synth}
Consider the transition system $G$ and its observer $\mc O(G)$ in \figr{f:synth}. The secret states are $X_S=\{2,5\}$ and the uncontrollable event $\Sigma_u=\{d\}$ (denoted by exclamation mark).
Both CSO and CSA result in the forbidden (non-safe) state set $\what X_f=\{\{2\},\{5\}\}$. Since the event $d$ is uncontrollable, the set of extended forbidden states in the observer $\what X^e_f=\{\{2\},\{3,4\},\{5\}\}$. Thus, the only remaining state in the supervisor $S=\mc O(G)_{\setminus \what X^e_f}$ is $\what X \setm \what X^e_f=\{\{0,1,2\}\}$, and the disabled events in this block state are $b$ and $c$, while the event $a$ is enabled.
\ebox
\e{example}

\subsection{Incremental supervisor generation by nonblocking preserving abstraction}
In \secr{s:extobs} it was shown how a forbidden state verification problem can be transformed to a nonblocking problem. More specifically, an observer $\mc O(G)$ has forbidden states if and only if the extended observer $\mc O_e(G)$ in \rf{eq:obse} has blocking states. To formulate this as a supervisor synthesis problem, the added $w$ self-loops in the local observers at the forbidden states are considered to be uncontrollable. This means that the transitions to the blocking states in the extended observer are uncontrollable. The source states at these uncontrollable transitions (the originally forbidden states), as well as the blocking states, will then be excluded in the nonblocking synthesis. 

Since the extended observer $\mc O_e(G)$ is modular, consisting of synchronized local extended observers, the same type of abstraction as in \secr{s:extobs} can be applied also before the nonblocking and controllable supervisor is computed. Instead of conflict equivalence that preserves nonblocking properties for verification purpose, a somewhat finer equivalence is required to be able to make synthesis on the abstracted observer. In \citep{flordal09, sahar14, sahar17}, a supervisor synthesis equivalence is proposed by which an incremental abstraction is performed similar to Algorithm 1 in \secr{handling}. This compositional synthesis, which is also implemented in the DES software tool Supremica \citep{olj:aff:sup:2006}, is applied to the modular extended observer $\mc O_e(G)$ in \rf{eq:obse}, including the uncontrollable events mentioned above.

\b{example}\label{ex:building:synth}
Consider again the $2$-story building with two elevators and the floor observers depicted in \figr{exf2e2}. The set of secret states for the floors are $X^{F^1}_S=\{3\}$ and $X^{F^2}_S=\{1,2'\}$. As can be seen in \figr{exf2e2}, the floor observers have no state exclusively including $3$ , while there is a state in the second floor observer that exclusively includes state $2'$. This state is a CSO non-safe state, which makes the second floor non-opaque. To make the whole system opaque, a supervisor $S$ is generated that restricts the second floor observer from entering the non-safe state. The resulting supervisor is shown in \figr{ex:sup}, where the synchronous composition $\mc O(F^2)\synch S$ does not include the non-safe state $2'$.

\tabr{tab:synth:trans} shows $|\what X|$, $|\what T|$ and the elapsed time $(t^S_e)$ for generating both compositional and monolithic supervisors, for the non-opaque building examples explained in Section~\ref{sec:ex}. The model has the same structure with similar set of secret states $X_S$ on each floor. The resulting computation times for the larger examples are about 3-6 times faster using the incremental synthesis procedure.
\hfill \ebox
\e{example}

\fig{t}{
\begin{tikzpicture}[xscale=1, yscale=1, automaton]
            \node[state,initial,initial text={$S$}] (1) at (-23em,-7em) { $1$};
            \node[state] (2) at (-17em,-7em) { $2$};
            \path[->]
              (1) edge [] node {$c^2_1,e^2_2$} (2)
              (1) edge [out=50,in=110, loop] node [above] {$e^2_1$} (1)
              (2) edge [out=50,in=110, loop] node [above] {$c^2_1,e^2_2$} (2);

\node[state,above,initial above,initial text={$\mathcal O(F^2)\synch S$},minimum size=1ex,rounded rectangle] (init) at (0em,-.25em)  {$\{1,2,3,4,2',4' \}$};
\node[state,minimum size=1ex,rounded rectangle]    (123)  at (9.5em,.5em)    {$\{1,2,3\}$};
\node at (4em,-1.4em) {\xspa{0.8em}$c^2_2$};
\node at (-5em,-1.4em) {\xspa{0.8em}$c^2_1$};
\node[state,minimum size=1ex,rounded rectangle]    (134)  at (-9.5em,.5em)   {$\{1,3,4\}$};
\node[state,minimum size=1ex,rounded rectangle]    (25)   at (4em,-5em)   {$\{2,2'\}$};
\node at (5.8em,-3.0em) {\xspa{0.8em}$e^2_1$};
\node[state,minimum size=1ex,rounded rectangle]    (46)   at (-4em,-5em)  {$\{4,4'\}$};
\node at (-6.5em,-3.0em) {\xspa{0.8em}$e^2_2$};
\node[state,minimum size=1ex,rounded rectangle]    (6)    at (-9.5em,-5em) {$\{4'\}$};
\node[state,minimum size=1ex,rounded rectangle]    (2)    at (9.5em,-10em)   {$\{2\}$};
\node[state,minimum size=1ex,rounded rectangle]    (4)    at (-9.5em,-10em)  {$\{4\}$};
\node[state,minimum size=1ex,rounded rectangle]    (13)   at (0em,-10em)  {$\{1,3\}$};
\path[->] 		
(init) edge [] node [above] {$c^2_1$} (123)
	   edge [] node [above] {$c^2_2$} (134)
       edge [] node [above] {\xspa{0.3em}$e^2_1$} (25)
       edge [] node [above] {$e^2_2$\xspa{0.5em}} (46)
(123) edge [out=-20,in=30, loop,distance=2.8em] node [right] {$c^2_1$} (123)
      edge [] node [below] {} (4)
(134) edge [out=200,in=150, loop,distance=2.8em] node [left] {$c^2_2$} (134)
      edge [] node [left] {$e^2_2$} (6)
      edge [] node [below] {} (2)
(6)  edge [] node [left] {$e^2_2$} (4)
(4)  edge [] node  {} (6)
(46) edge [out=210,in=150,loop,distance=2.8em] node  {} (46)
     edge [] node [below]  {$c^2_2$\xspa{0.3em}} (13)
(25) edge [out=-30,in=30,loop,distance=2.8em] node  {} (25)
     edge [] node [below] {$c^2_1$} (13)
(4) edge [] node  {} (13)
(2) edge [] node  {$c^2_1$} (13)
(13) edge [] node  {} (2)
(13) edge [] node  {$c^2_2$} (4);
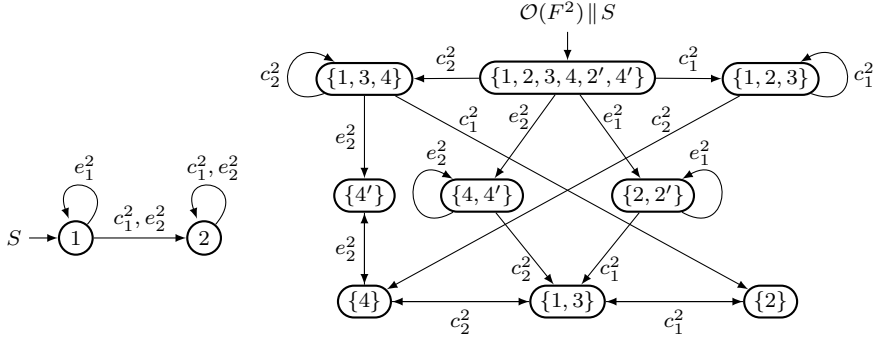
\end{tikzpicture}}{Supervisor $S$ for the whole system, and the observer of the second floor $\mc O(F^2)$ after synchronization with the supervisor.}{ex:sup}


\tab{b}{Number of states and transitions of supervisors based on incremental and monolithic synthesis for non-opaque buildings with $n$ floors and $m$ elevators. The larger examples obtain modular supervisors based on incremental synthesis. \yspa{2ex}\hfill}
{\begin{tabular}{c|ccc|ccc} \toprule
 $F^i/E^j$-model & \multicolumn{3}{ c  }{$S$ incremental synth. \yspatb{0.8ex}{1ex}}& \multicolumn{3}{ c }{$S$ monolithic synth. \yspatb{0.8ex}{1ex}}  \\ \hline
$(n,m)$ & $|\what X|$ & $|\what T|$ & $t^{S}_e$ (s) \yspatb{0.6ex}{1ex}& $|\what X|$ & $|\what T|$ & $t^{S}_e$ (s) \yspatb{0.6ex}{1ex} \\ \midrule
(1,3)\yspatb{0.2ex}{1.5ex}  & 1 & 0 & 0.02 & 1&0&0.02\\
(2,2)\yspa{1.5ex}  & 2 & 5 & 0.05 & 2&5& 0.13\\
(2,3)\yspa{1.5ex}  & 3 & 10 & 0.09 & 3 & 6 & 0.31  \\
(3,3)\yspa{1.5ex}  & (3,7) & (10,66) & 70 &61&700& 441 \\
(3,4)\yspa{1.4ex}  & (3,6) & (12,60) & 54 & -- & -- &-- \\ \bottomrule
\end{tabular}}{tab:synth:trans}

\section{Conclusions}
To tackle the exponential observer generation complexity for current-state opacity/ anonymity verification and enforcement, but also the complexity that arises when modular subsystems are synchronized, an incremental local observer abstraction is proposed. The two notions of current-state opacity and current-state anonymity are formulated based on state labels in transition systems that are naturally extended to modular systems. Non-safe states in corresponding local observers are then considered as forbidden states. By introducing simple detector automata, this problem is easily transformed to a nonblocking problem for which efficient existing abstraction methods are applied. This transformation to a nonblocking problem by detector automata is a generic technique with a great potential. A recent alternative example is incremental abstraction for verification of diagnosability \citep{mona_diag:2019}. 

The main theoretical development in this paper is the combined incremental abstraction and observer generation for systems with shared unobservable events. Due to the need for additional temporary state labels, the more general but less efficient visible bisimulation is then used as abstraction. An interesting alternative is to introduce arbitrary state labels in the more efficient abstractions that also have been used in this paper, but then only applied to transition systems without shared unobservable events. The efficiency of the proposed methods is demonstrated through a modular multiple floor/elevator building example. Both verification and supervisor synthesis to enforce a secure system are  investigated. The results show a great improvement when abstraction is included especially for verification, while there is hope for some further improvements in the observer based supervisor synthesis.

\bibliographystyle{spbasic} \bibliography{Automation-mona} \end{document}